\documentclass[aps,prl,twocolumn,superscriptaddress,longbibliography]{revtex4-2}

\usepackage{graphicx}
\usepackage{dcolumn}   % needed for some tables
\usepackage{bm}        % for math
\usepackage{amssymb}   % for math
\usepackage{amsmath}
\usepackage{url}
\usepackage{layout}
\usepackage{color}
\usepackage{epsfig}
\usepackage{graphicx}
\usepackage{mathrsfs}\usepackage{bm}\usepackage{appendix}\usepackage{color}

\usepackage{multirow}
\usepackage[thinlines]{easytable}
\usepackage{makecell}
\usepackage{tikz,pgf}
\usepackage{booktabs}
\usepackage{float}
\usepackage{subfigure}
\usepackage[normalem]{ulem}

\usepackage{cancel}

\usepackage{hyperref}
\hypersetup{colorlinks,allcolors=blue}

\newcommand{\rme}{\mathrm{e}}

\begin{document}

% ----- title [begin] -----
\title{N\'eel Ordered Magnetic Phases in Bipartite Quasicrystals}
% ----- title [end] -----

% ----- author [begin] -----
\author{Jia-Heng Ji}
\thanks{These authors contributed equally to this work.}
\affiliation{School of Physics, Beijing Institute of Technology, Beijing 100081, China}
\author{Zhi-Yan Shao}
\thanks{These authors contributed equally to this work.}
\affiliation{School of Physics, Beijing Institute of Technology, Beijing 100081, China}
\author{Yu-Bo Liu}
\thanks{These authors contributed equally to this work.}
\affiliation{Institute of Theoretical Physics, Chinese Academy of Sciences, Beijing 100190, China}
\author{Wei Li}
\email{w.li@itp.ac.cn}
\affiliation{Institute of Theoretical Physics, Chinese Academy of Sciences, Beijing 100190, China}
\author{Fan Yang}
\email{yangfan\_blg@bit.edu.cn}
\affiliation{School of Physics, Beijing Institute of Technology, Beijing 100081, China}
% ----- author [end] -----

% ----- abstract [begin] -----
\begin{abstract}
Magnetism is a fundamental research area in which the recently proposed altermagnetism (AM) has become an emergent frontier. 
Very recently, the quasicrystal (QC) was proposed as a possible platform to realize AM. 
However, the existence of AM in QCs still lacks vigorous evidence. 
In this work, we adopt the sign-problem-free projector quantum Monte Carlo (PQMC) algorithm to investigate the magnetic phases in the half-filled Hubbard models in various 2D bipartite QCs, and always obtain N\'eel ordered states. 
While the N\'eel states in bipartite crystals are usually antiferromagnetism (AFM), we find it common that those in bipartite QCs can also be AM or ferrimagnetism (FIM). 
Based on symmetry analysis, combined with our comprehensive PQMC results, we propose a general criterion for determining the magnetism classes of the N\'eel states in a bipartite QC: 
According to whether the two sublattices are related by inversion, a non-inversion point-group operation such as mirror or rotation, or no operation about the unique symmetry center, the corresponding N\'eel state is AFM, AM or FIM / geometry-enforced AM (G-AM), respectively.
The G-AM is a special type of $s$-wave AM caused by the unique QC structure formed through projection from high-dimensional crystals to lower dimensions, which is rare in crystals.
For example, our results yield symmetry-protected AM for the two $D_4$-symmetric Thue-Morse QCs and G-AM for the $D_5$-symmetric Penrose QC at half-filling. 
Our results provide a solid foundation for experimental investigations and potential applications of different classes of magnetism in QCs. 
\end{abstract}
% ----- abstract [begin] -----

\maketitle

%%%%% Main Text %%%%%
% ----- Introduction [begin] -----
\paragraph{\textcolor{blue}{Introduction. ---}}
Magnetism plays a crucial role in condensed matter physics, with ferromagnetism (FM) and antiferromagnetism (AFM) representing two traditional magnetic phases.
The representative physical origins of FM include the Stoner criterion\cite{stoner1938Stoner} that considers direct exchange interactions between itinerant electrons and the double-exchange mechanism\cite{zener1951doubleexchange, gennes1960doubleexchange} that describes interactions of itinerant electrons with localized spins. In contrast, AFM usually emerges through Fermi surface nesting\cite{fawcett1988nesting, scalapino1995nesting, imada1998nesting} in weak-coupling systems or superexchange interactions\cite{kramers1934superexchange, anderson1950superexchange, goodenough1958superexchange, kanamori1959superexchange} in strong-coupling systems.
Composite lattice structures can further induce ferrimagnetism (FIM) by the superexchange mechanism\cite{goodenough1955superexchange}, which is a subtype of FM with opposite magnetic moments of unequal magnitude on distinct sublattices yielding residual net magnetization.

Recently, altermagnetism (AM) has been considered as a distinct class of magnetic phase\cite{naka2019spin, ahn2019antiferromagnetism, hayami2019momentum, vsmejkal2020crystal, yuan2020giant, shao2021spin, mazin2021prediction, ma2021multifunctional, yuan2021prm, zhou2024crystal, zhang2024gate, bai2024altermagnetism, han2024electrical, zhou2025manipulation}, characterized by zero net magnetism similar to AFM and spin-split electronic band structures similar to FM. 
Theoretically, AM can be realized by various mechanisms, e.g. the presence of nonmagnetic atoms\cite{fender2025altermagnetism, vsmejkal2022beyond}, spontaneously broken orbital order\cite{leeb2024spontaneous, vila2025orbital, meier2025anti}, Pomeranchuk instabilities\cite{wu2004dynamic, wu2007fermi, jungwirth2025altermagnetism, qian2025fragile} or spin-cluster engineering\cite{zhu2025design}.
Currently, although there exist experimental evidences to support that some materials are possible candidates for AM~\cite{krempasky2024altermagnetic, olena2024obsevation, lee2024broken, ding2024large, liu2024absence, lin2024observation, jiang2025metallic, osumi2024observation, graham2025local, reichlova2024observation, reimers2024direct, yang2025three, urru2025Gtype}, the confirmed ones among them are still rare.

To explore more possibilities to realize AM, researchers recently considered AM in quasicrystals (QCs) \cite{chen2025quasi, li2025quasi, shao2025classification}.
The QC is an aperiodic structure that possesses long-range-order, in which the higher-fold rotational symmetries and the absence of translational symmetry\cite{levine1984quasi, levine1986quasi, socolar1986quasi} produce various novel phases\cite{Wessel2003, Kraus2012, Thiem2015, Andrade2015, Otsuki2016, Koga2017Penrose, Huang2018, Huang2019, Longhi2019, Koga2020AB, Miyazaki2020, caoye2020, Duncan2020, Manna2024, Hua2021, Peng2021, Ghadimi2021, Akihisa2021, Wang2022, Koga2022, Ghosh2023, Chen2023, Yu_Bo_Liu2023, Uri2023, Araujo2024, Yang2024, Yu_Bo_Liu2024, LiuLu2025, chen2025quasi, li2025quasi, AJagannathan1, AJagannathan2, AJagannathan3, Uri2025, chen2025quasi, li2025quasi, shao2025classification}, including the magnetic phases\cite{Wessel2003, Thiem2015, Otsuki2016, Koga2017Penrose, Koga2020AB, Koga2022, Ghosh2023, AJagannathan1, chen2025quasi, li2025quasi, shao2025classification}.
In Refs.\cite{chen2025quasi,li2025quasi}, the possibility of realizing AM in some 2D QCs is proposed through mean-field (MF) studies. In Ref.\cite{shao2025classification}, it is pointed out that without the translational symmetry to recover the broken parity ($\mathcal{P}$) and time-reversal ($\mathcal{T}$) symmetry, point-group symmetries other than $\mathcal{P}$ can protect AM, making AM more common in QCs than in crystals.
Furthermore, since a QC can host at most one inversion center, magnetic phases can be easily classified according to the irreducible representations (IRRPs) of the point-group\cite{shao2025classification}. 
Ref.\cite{shao2025classification} points out that nearly half of the 1D IRRPs correspond to AM, implying the widespread presence of AM in QCs.

However, the existence of AM in QCs still lacks rigorous evidence.
On the one hand, the possible realization of AM in QCs investigated by self-consistent MF approach is not persuasive:
Firstly, the MF theory cannot capture the quantum fluctuations.
Secondly, due to absence of translational symmetry, it is difficult to determine the spatial distribution pattern of the magnetic moments in QCs, easily leading to being trapped at local energy minima.
On the other hand, although the classification\cite{shao2025classification} indicates which IRRPs correspond to AM, it is not easy to determine which IRRP the ground state of a given QC belongs to.
To address these problems, we employ the sign-problem-free projector quantum Monte Carlo (PQMC) algorithm\cite{sugiyama1986dqmc, sorella1989dqmc, sorella1988dqmc} to study the magnetic phases in bipartite QCs by solving the half-filled large-$U$ Hubbard model, which turn out to be N\'eel-ordered states. 
Intriguingly, in contrast to the cases in bipartite crystals wherein the N\'eel states are usually AFM, the N\'eel states in QCs can be AFM, AM or FIM, determined by symmetries and geometries of the QCs.

\begin{figure*}[tbp]
    \centering
    \includegraphics[width=0.8\linewidth]{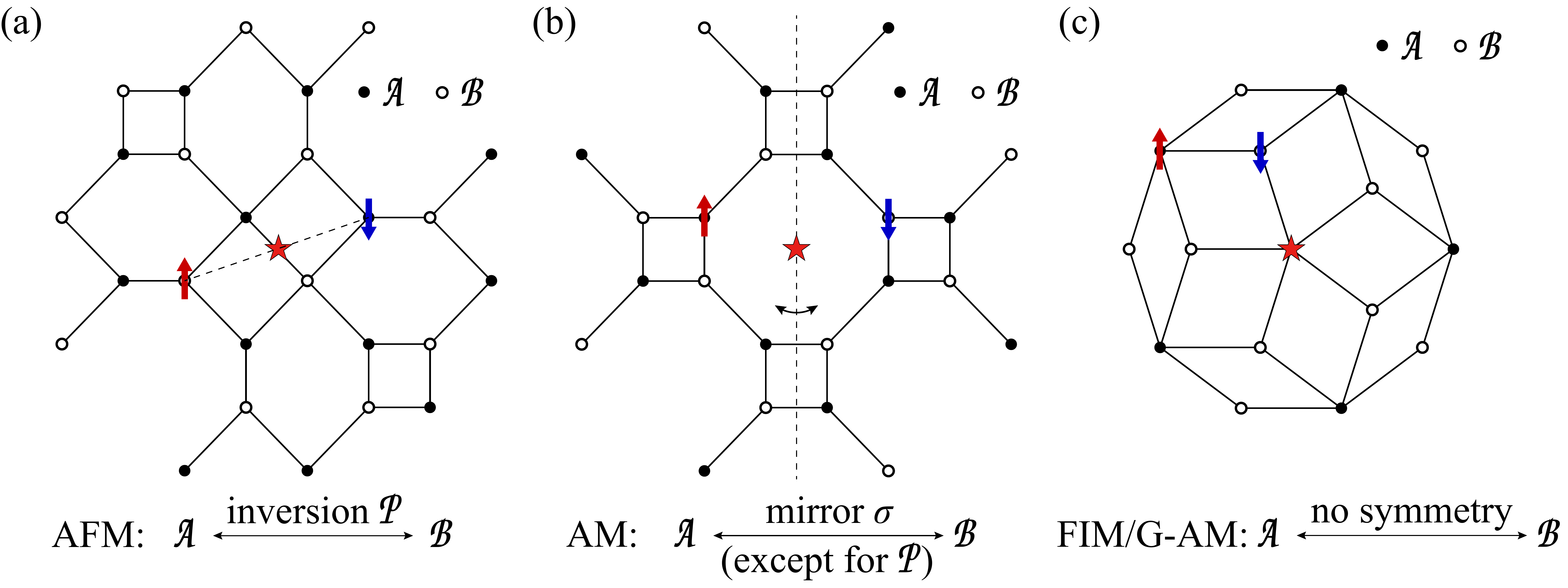}
    \caption{Schematic diagrams of the lattice geometry-based criterion for determining the magnetic classes of N\'eel ordered states in bipartite QCs.
    (a) N\'eel ordered AFM is allowed in, e.g. the $D_2$-symmetric Thue-Morse QC, whose two sublattices are related by $\mathcal{P}$. 
    A pair of sites related by $\mathcal{P}$ (dashed line) with opposite magnetic moments (red up arrow for spin-up and blue down arrow for spin-down) are marked in (a).
    (b) N\'eel ordered AM emerges in, e.g. the $D_4$-symmetric Thue-Morse QC, where a non-inversion point-group operation (e.g., the mirror $\sigma$) instead of $\mathcal{P}$ connects the two sublattices. The dashed line represents a mirror axis. The red (blue) arrow indicates the same meaning as in (a). 
    (c) N\'eel ordered FIM / G-AM is present in, e.g. the $D_5$-symmetric Penrose QC, in which no symmetry operations relate the two sublattices.
    Within this class, when the two sublattices host unequal (equal) site numbers, the magnetic class is FIM (G-AM).
    In (a-c), the red stars label the symmetry centers, and the solid (hollow) circles denote the sites in sublattice $\mathcal{A}$ ($\mathcal{B}$) and the solid lines represent the NN or NNN bonds.}
    \label{fig:criterion}
\end{figure*}

In this work, we investigate the magnetic phases in the half-filled large-$U$ Hubbard models in various 2D bipartite QCs, which are expected to be N\'eel ordered phases. 
We propose a general criterion for determining the magnetism classes of the N\'eel states in these bipartite QCs: 
Due to lack of translational symmetry, there exists one unique point-group symmetry center in a QC. 
According to whether the two sublattices are related by the inversion operation $\mathcal{P}$, a non-inversion point-group operation such as a mirror or rotation, or no operation about this symmetry center, the corresponding N\'eel state is AFM, AM or FIM / geometry-enforced AM (G-AM), respectively. 
For verification, we solve the ground state via the sign-problem-free PQMC algorithm to determine the magnetic pattern and study the spectroscopic properties. 
The PQMC simulation verifies our criterion and reveals the robust existence of symmetry-protected AM, AFM and G-AM N\'eel orders in bipartite QCs, providing a theoretical foundation for future experimental investigations and potential applications.
% ----- Introduction [end] -----

% ----- Model [begin] -----
\paragraph{\textcolor{blue}{Model and Approach. ---}}
As the beginning of our study, we construct the Hubbard model in QCs as
\begin{equation}\label{eq:H_Hubbard}
    H = H_\text{TB} + U \sum_{i} n_{i\uparrow} n_{i\downarrow},
\end{equation}
where $n_{i\sigma}$ is the particle number of spin-$\sigma$ at site $i$ and $H_\text{TB}$ represents the tight-binding (TB) Hamiltonian, including up to the nearest neighbor (NN) or the next nearest neighbor (NNN) inter-sublattice hoppings. 
See Supplementary Materials (SM)\cite{SM} for details.
We set the hopping integral along the NN bonds $t_1$ as an energy unit and $U = 4t_1$ throughout this work.
In this parameter regime, the half-filled bipartite Hubbard models studied here are in the insulating Mott regime, and our work focuses on the magnetic classes of these insulating N\'eel ordered phases.

We solve the Hamiltonian (\ref{eq:H_Hubbard}) using the sign-problem-free PQMC algorithm with a non-uniform time-slice strategy for half-filled bipartite QCs.
More details are provided in the SM~\cite{SM}.

\paragraph{\textcolor{blue}{Magnetic classes of N\'eel orders in bipartite QCs. ---}} 
For large-$U$, the two sublattices within a bipartite lattice tend to host magnetic moments with opposite orientations, which constitute the N\'eel order, so that the energy of the low-energy effective superexchange interaction between the two sublattices is minimized. 
However, this does not ensure a zero net magnetism in QCs: 
Firstly, the inevitable quantum fluctuations lead to partial polarization in each sublattice. 
Secondly, different sites within a QC reside in distinct local environments and experience different quantum fluctuation strengths, and thus host different magnitudes of magnetic moment, unless they are related by any point-group operation. 
Therefore, in a bipartite QC, based on whether the two sublattices are related by any point-group symmetry element and by which symmetry, there exist three possibilities.

The first is the N\'eel ordered AFM. 
If the two sublattices are related by $\mathcal{P}$, as shown in Fig. \ref{fig:criterion} (a), each pair of sites related by $\mathcal{P}$ must belong to different sublattices, and will thus accommodate magnetic moments with equal magnitudes but opposite orientations. 
Therefore, the global magnetic pattern will be flipped after $\mathcal{P}$, which can be recovered after an extra $\mathcal{T}$ operation, leading to the combined $\mathcal{PT}$ symmetric AFM phase.

The second is the N\'eel ordered AM. 
If the two sublattices are not related by $\mathcal{P}$, but are related by an alternative point-group symmetry, e.g. the mirror $\sigma$, as illustrated in Fig. \ref{fig:criterion} (b), this symmetry can also ensure zero net magnetism. 
In this case, the N\'eel state lacks the $\mathcal{PT}$ symmetry, leading to a N\'eel ordered AM. 
Note that this differs from the situation in crystals. 
For example, in a square lattice, although the two sublattices cannot be related by a plaque-centered $\mathcal{P}$ operation, like that illustrated in Fig. \ref{fig:criterion} (b), they can alternatively be related by a bond-centered $\mathcal{P}$ operation, leading to a $\mathcal{PT}$ symmetric AFM N\'eel state. 
However, in QCs, due to the absence of translational symmetry, there can exist at most one inversion center~\cite{shao2025classification} as marked in Fig. \ref{fig:criterion}(b), which relates one sublattice to itself by $\mathcal{P}$, leading to the $\mathcal{PT}$ symmetry broken N\'eel ordered AM phase. 

The third is the N\'eel ordered FIM / G-AM. 
If there is no point-group symmetry that relates the two sublattices, as shown in Fig.\ref{fig:criterion} (c), any site in one sublattice transfers to another site in the same sublattice after any symmetry operation (e.g., $\mathcal{P}$, $\sigma$ or rotation). 
In this case, no symmetry ensures the zero net magnetism. 
However, Lieb's theorem\cite{LiebTheorem} states that, for a repulsive Hubbard model on a bipartite lattice at half filling, the total spin of the ground state is $S=\frac{1}{2}|N_{\mathcal A}-N_{\mathcal B}|$, where $\mathcal A$ and $\mathcal B$ are the two sublattices of the bipartite lattice.
Therefore, if the two sublattices have unequal site densities, the N\'eel state is FIM; if their site densities are equal, the N\'eel state is G-AM.

{\bf In summary, we have: In a bipartite QC, depending on whether the two sublattices are related by the $\mathcal{P}$ operation, a non-inversion point-group operation such as mirror or rotation, or no symmetry operation about the unique symmetry center, the corresponding N\'eel order is AFM, AM, or FIM / G-AM, respectively.}

To verify this proposal, we employ the sign-problem-free PQMC approach to study the half-filled large-$U$ Hubbard models in various bipartite QCs as examples.
% ----- Criterion [end] -----

\begin{figure}[tbp]
    \centering
    \includegraphics[width=1\linewidth]{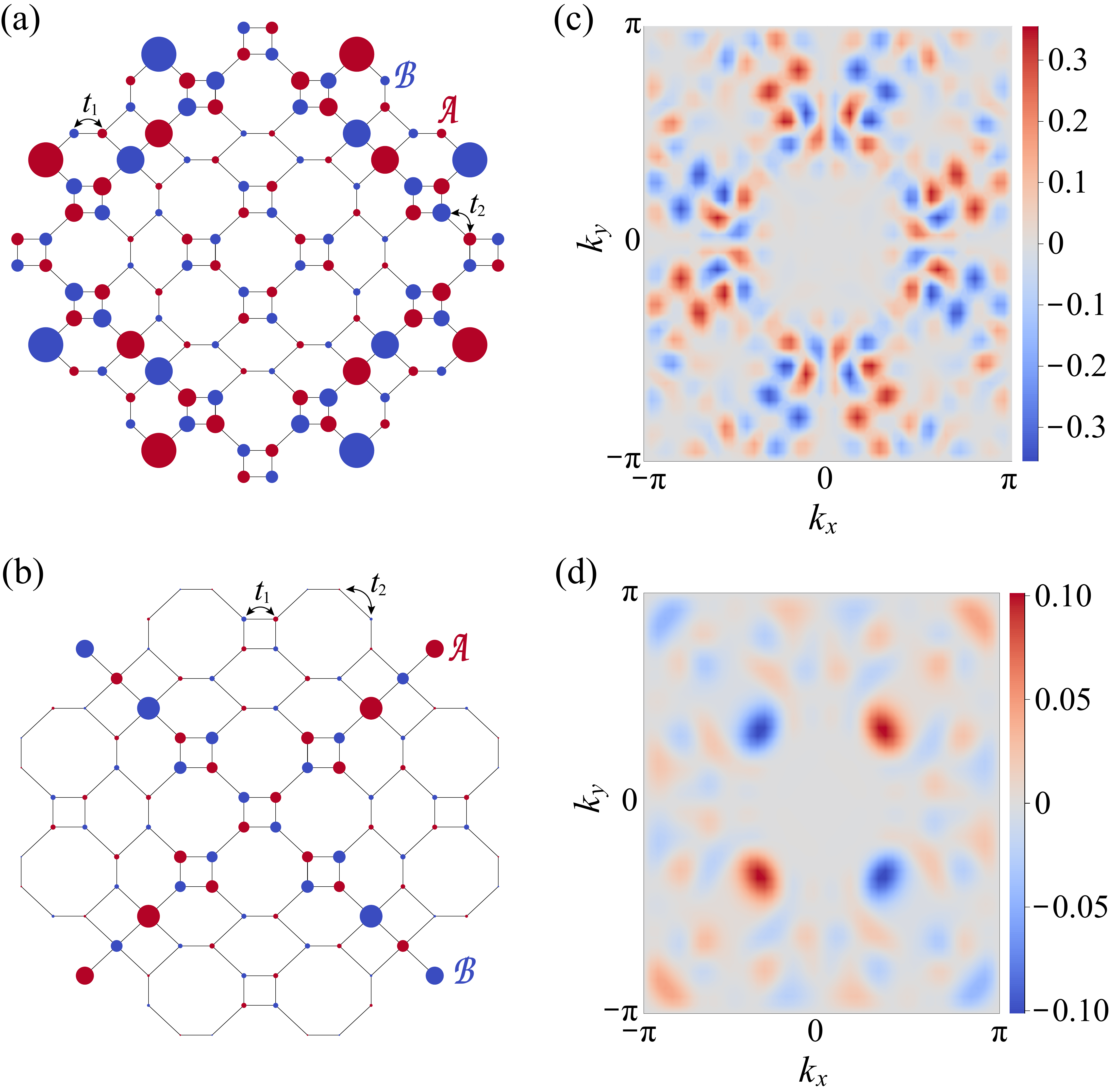}
    \caption{
    (a, b) The distribution of the magnetic moments in the $D_4$-symmetric Thue-Morse lattice hosting (a) 120 sites  with the central 8-site cluster, and (b) 104 sites with the central 4-site cluster. In (a-b), the red (blue) dots represent spin-$\uparrow$ ($\downarrow$), and the size of each dot represents magnitude of the magnetic moment. The central black stars label the symmetry center and the solid lines with $t_1$ and $t_2$ represent the NN and NNN hoppings.
    (c, d) The spin-resolved spectral density difference $\mathcal{A}_\uparrow(\bm{p},\omega) - \mathcal{A}_\downarrow(\bm{p},\omega)$ as a function of momentum $\bm{p}$ for (c) the lattice exhibited in (a) with $\omega=-0.33t_1$, and (d) the lattice exhibited in (b) with $\omega=-0.2t_1$.}
    \label{fig:AM_result}
\end{figure}

% ----- Examples [begin] -----
\paragraph{\textcolor{blue}{Examples. ---}}
% --- AM ---
Firstly, we explore the possibility of N\'eel ordered AM in bipartite QCs.
According to the above proposal, this requires that the two sublattices are connected by point-group operations other than $\mathcal{P}$.
The two different $D_4$-symmetric Thue-Morse lattices shown in Figs. \ref{fig:AM_result} (a) and (b) satisfy this condition: In both QCs, the two sublattices are related by mirror or rotation operation instead of $\mathcal{P}$. In the following, we verify the existence of N\'eel ordered AM in the two QCs by the sign-problem-free PQMC approach.

The TB Hamiltonians on the two Thue-Morse lattices include hopping terms along the NN- ($t_1$) and the NNN ($t_2$) bonds, as indicated in Figs. \ref{fig:AM_result} (a) and (b) by the solid lines. 
The magnetic patterns on the two lattices obtained by PQMC are shown in Figs. \ref{fig:AM_result} (a) and (b), which illustrate the N\'eel orders with opposite spin orientations in the two sublattices. 
The magnetic pattern shown in Fig. \ref{fig:AM_result} (a) exhibits a $d_{x^2-y^2}\cdot d_{xy}$-wave symmetry under the $D_4$ point-group around the symmetry center marked by the solid pentagram: It is unchanged under $\mathcal{P}$ or the rotation $c^1_4$ and changes sign under the mirror reflection about the four axes of symmetry. 
The magnetic pattern shown in Fig. \ref{fig:AM_result} (b) exhibits a $d_{xy}$-wave symmetry: It is unchanged under $\mathcal{P}$ or mirror reflection about the diagonal directions and changes signs under $c^1_4$ or mirror reflection about the $x$- or $y$-axes.

The above two N\'eel states shown in Fig. \ref{fig:AM_result} remain unchanged under $\mathcal{P}$ and then flip spins after an extra $\mathcal{T}$ operation, which break the $\mathcal{PT}$ symmetry and are thus AM. To further characterize AM in QCs, we calculate the spin-resolved angle-resolved photoemission spectrum $\mathcal{A}_{\uparrow,\downarrow}(\bm{p},\omega)$\cite{chen2025quasi,li2025quasi,dornellas2025alter} (see SM\cite{SM} for details), which indicates the spectrum weight for spin-$\uparrow$ or $\downarrow$ at a certain energy level $\omega$ to detect momentum $\bm{p}$. As shown in Figs. \ref{fig:AM_result} (c) and (d), the spectrum differences $\mathcal{A}_\uparrow(\bm{p},\omega) - \mathcal{A}_\downarrow(\bm{p},\omega)$ as functions of $\bm{p}$ exhibit non-zero value with corresponding symmetries, characterizing AM.

These examples demonstrate the great potential of the QC as a platform to realize AM. Note that the magnetic pattern shown in Fig. \ref{fig:AM_result} (b) exhibits the same $d_{xy}$-wave symmetry as that in the square-lattice N\'eel state. 
In the latter state, the plaque-centered $\mathcal{PT}$ symmetry is also broken, like that illustrated in Fig. \ref{fig:AM_result} (b). 
However, the broken $\mathcal{PT}$ symmetry can be recovered after a subsequent translation by a unit vector. Here in the QC, the lack of translational symmetry prevents recovery of the $\mathcal{PT}$ symmetry, making AM more easily realized.

\begin{figure}[tbp]
    \centering
    \includegraphics[width=1\linewidth]{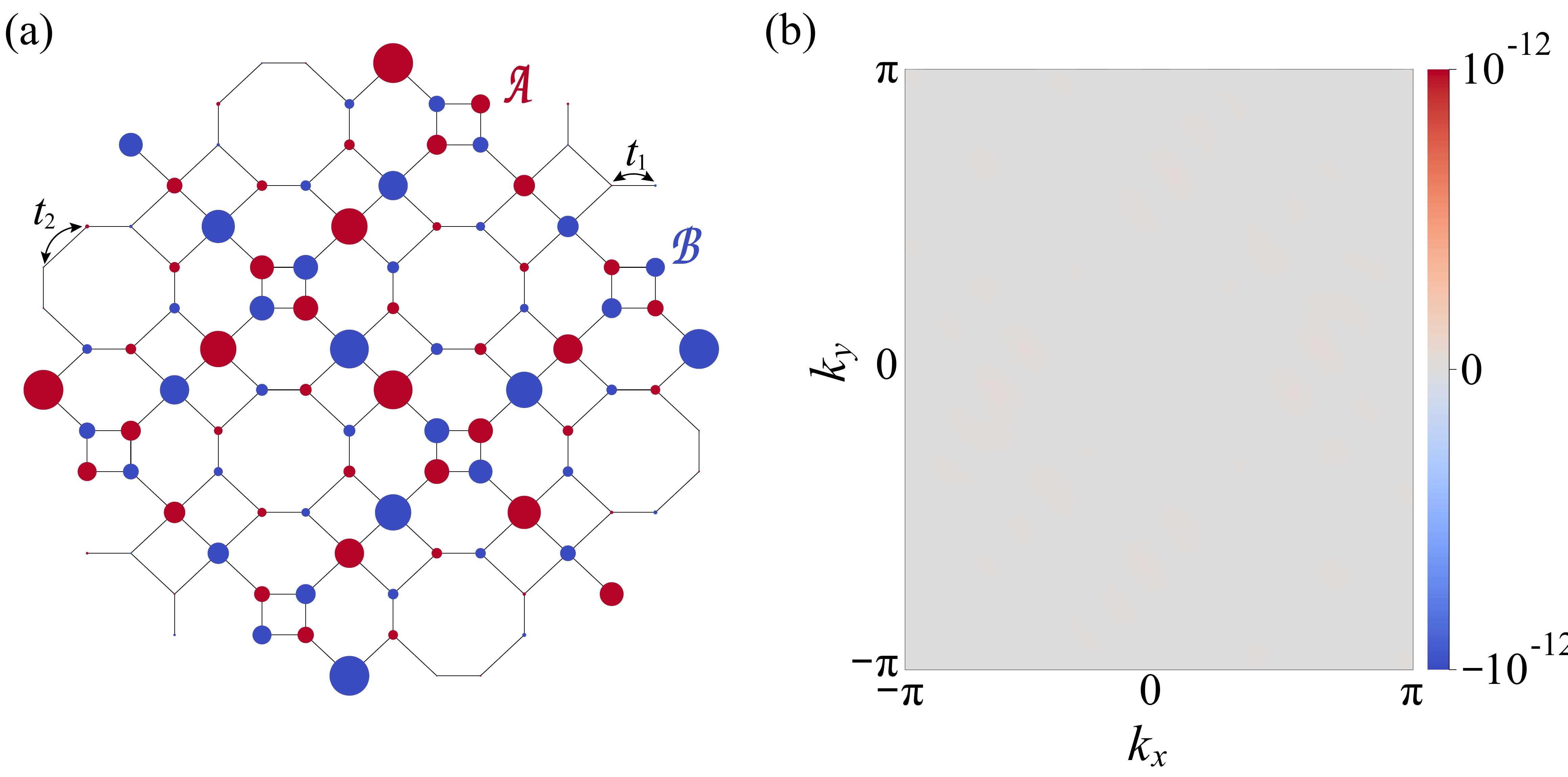}
    \caption{
    (a) The distribution of the magnetic moment in the 104-site $D_2$-symmetric Thue-Morse lattice. 
    (b) The spin-resolved spectral density difference $\mathcal{A}_{\uparrow}(\bm{p},\omega) - \mathcal{A}_{\downarrow}(\bm{p},\omega)$ as a function of momentum $\bm{p}$ for arbitrary $\omega$.}
    \label{fig:AFM_result}
\end{figure}

% --- AFM ---
Secondly, we consider the possibility of N\'eel ordered AFM in biparitite QCs. According to our proposal, this requires that the two sublattices are connected by $\mathcal{P}$. A typical example is the $D_2$-symmetric Thue-Morse lattice shown in Fig. \ref{fig:AFM_result} (a), in which the two sublattices are indeed related by $\mathcal{P}$. The magnetic pattern on this lattice obtained by our PQMC calculation is shown in Fig. \ref{fig:AFM_result} (a), which exhibits a N\'eel order with opposite spin orientations in the two sublattces. This pattern exhibits the $p_{x-y}$-wave symmetry under the $D_2$ point-group: It is unchanged under the mirror reflection about the $x+y=0$ line and changes sign under $\mathcal{P}$, $c^1_2$ rotation or mirror reflection about the $x-y=0$ line. Obviously, such a magnetic pattern would be flipped under $\mathcal{P}$ and recovered after a subsequent $\mathcal{T}$ operation, belonging to the $\mathcal{PT}$-symmetric AFM phase. The spectrum difference $\mathcal{A}_{\uparrow}(\bm{p},\omega) - \mathcal{A}_{\downarrow}(\bm{p},\omega)$ shown in Fig. \ref{fig:AFM_result} (b) illustrates zero value for arbitrary $(\bm{p},\omega)$, characterizing AFM.

\begin{figure}[tbp]
    \centering
    \includegraphics[width=1\linewidth]{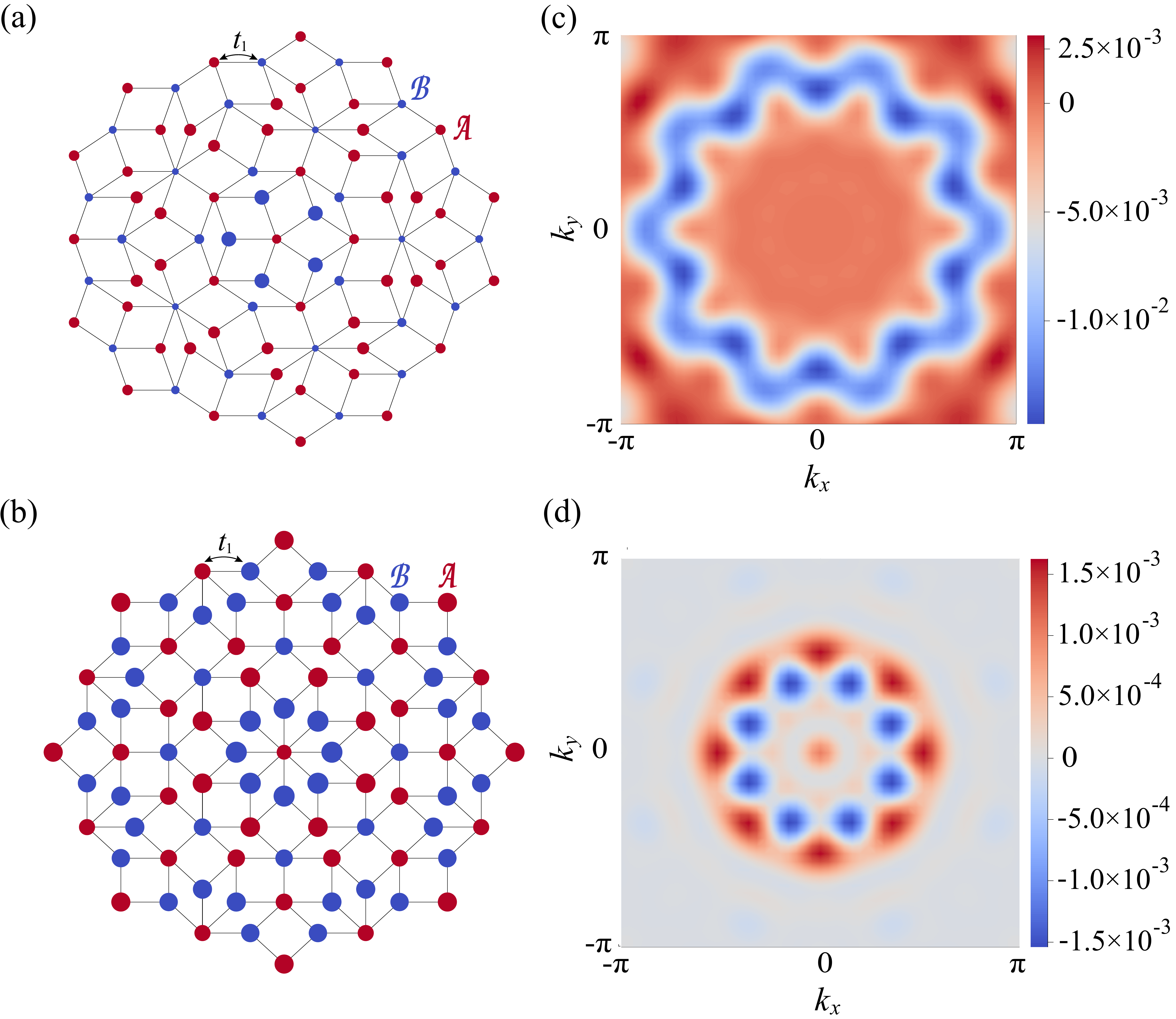}
    \caption{
    (a, b) The distribution of the magnetic moments in (a) the 86-site $D_5$-symmetric Penrose lattice and (b) the 89-site $D_8$-symmetric Ammann-Beenker lattice.
    (c, d) The corresponding spin-resolved spectral density difference $\mathcal{A}_\uparrow(\bm{p},\omega) - \mathcal{A}_\downarrow(\bm{p},\omega)$ as a function of momentum $\bm{p}$ for the lattices shown in (a) with $\omega=-0.44t_1$ and (b) with $\omega=-2.25t_1$, respectively.}
    \label{fig:FM_result}
\end{figure}

% --- FM ---
Finally, we study the possibility of N\'eel ordered FIM / G-AM in bipartite QCs.
According to our proposal, this requires that the two sublattices cannot be related by any point-group operation.
Within this symmetry-based class, unequal sublattice site densities lead to FIM, whereas equal sublattice site densities give G-AM for the half-filled bipartite Hubbard model.
A class of bipartite lattices which satisfy this condition is those with a single central site as the symmetry center, which is never related to another site by any point-group operation. 
The $D_5$-symmetric Penrose lattice and the $D_8$-symmetric Ammann-Beenker lattice shown in Figs.~\ref{fig:FM_result}(a) and (b) belong to this class, for which the TB Hamiltonians include only the NN hoppings indicated by the solid lines.

In the thermodynamic limit, the two sublattices have equal site densities in both QCs \cite{Koga2017Penrose,Koga2020AB}.
For the half-filled Hubbard model on a bipartite lattice, Lieb's theorem\cite{LiebTheorem} states that the total spin of the ground state is $S=\frac{1}{2}|N_{\mathcal A}-N_{\mathcal B}|$.
When the total spin $S$ is nonzero, the total magnetic moment of the ground state is at least $N \times S$, where $N=N_{\mathcal A}+N_{\mathcal B}$ is the total number of sites.
Therefore, the ground states on these two QCs have zero net magnetism in the thermodynamic limit.

The magnetic patterns obtained from our PQMC calculations in Figs.~\ref{fig:FM_result}(a) and (b) display N\'eel order with $s$-wave symmetry, remaining unchanged under all point-group operations.
Since these real-space patterns are not invariant under the combined operation of $\mathcal{T}$ and any $D_5$ or $D_8$ point-group operation, the zero net magnetism and the absence of combined $\mathcal{PT}$ symmetry together identify a geometry-enforced $s$-wave AM.
Figs.~\ref{fig:FM_result}(c) and (d) show the spin-resolved spectral density differences for the two G-AM states. They remain nonzero, momentum-dependent, and exhibit an $s$-wave pattern. 

Intriguingly, the $s$-wave G-AM is distinct from such higher-symmetry forms as the $d$-, $g$-wave usually associated with symmetry-protected AM.
Such an $s$-wave G-AM is absent or rare in crystals and originates from the higher-to-lower-dimensional projection construction of QCs~\cite{Koga2017Penrose, Koga2020AB}, where the sublattice site-density equality can be regarded as a remnant of the higher-dimensional parent-lattice symmetry. 
Therefore, G-AM represents a new type of AM unique to QCs.
% ----- Examples [end] -----

% ----- Conclusion [begin] -----
\paragraph{\textcolor{blue}{Conclusion and Discussion. ---}}
We have investigated the magnetic phases in half-filled large-$U$ Hubbard models in various bipartite QCs, demonstrating the feasibility of QCs as a platform to realize AFM, symmetry-protected AM, and FIM / G-AM. 
Based on symmetry analysis, combined with our comprehensive PQMC results, we propose a general criterion to determine the magnetism classes of the N\'eel states in bipartite QCs, which brings convenience in the exploration and application of magnetism in QCs.

This work focuses on half-filled Hubbard models with fixed $U$ in bipartite QCs.
With tuning $U$ and the doping, more intriguing phases are expected. 
On the one hand, in the case of weak $U$, without driving by Fermi surface nesting in QCs, it is fascinating to figure out the emergent magnetic patterns. 
On the other hand, upon doping, unconventional superconductivity (SC) could emerge. 
Particularly, the AM or FIM fluctuations might mediate triplet SC\cite{zhu2023topological, parthenios2025spin, wu2025intra, ma2025possible, lu2025inter, bao2015superconductivity, zhou2017theory, wu2015triplet, chen2018progress, jie2021spin, sheng2019nearly, jiao2020chiral, hayes2021multicomponent, aoki2022unconventional, lewin2023review, zhang2022kondo, squire2023superconductivity, shan2025emergent}. These topics are left for future study.
% ----- Conclusion [end] -----

% ----- Acknowledgement [begin] -----
~~~~
\noindent{{\bf Acknowledgment}}
We are grateful to the stimulating discussions with Cong Zhang.
F.Y. is supported by the national natural science foundation of China under the Grant Nos. 12574141, 12234016, 12074031. Y.-B. L. is supported by the postdoctoral innovation talent support program (Grant No. A5B10039).
~~~~~
% ----- Acknowledgement [end] -----

\bibliography{reference}

%%%%% Supplemental Material %%%%%
\appendix
\onecolumngrid
\section*{Supplementary Materials}

\renewcommand{\theequation}{S\arabic{equation}}
\renewcommand{\thefigure}{A\arabic{figure}}
\setcounter{equation}{0}
\setcounter{figure}{0}

% ----- Sec A [begin] -----
\section{A. Hubbard Model in Different Bipartite Quasicrystals}
\label{appendix_A}

\begin{figure}[htbp]
    \centering
    \includegraphics[width=0.99\linewidth]{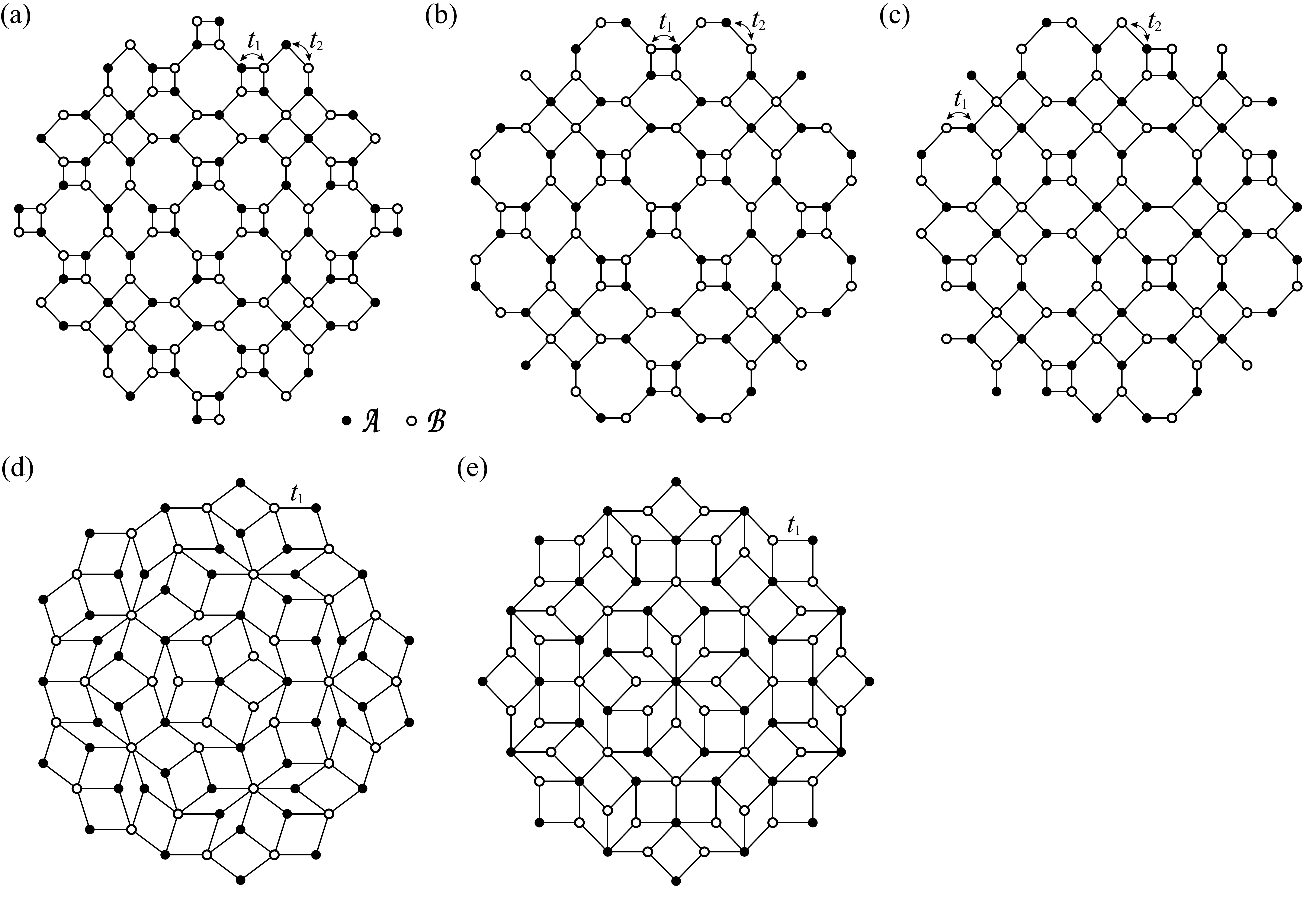}
    \caption{
    (a-e) The schematic diagram of $D_4$ symmetric Thue-Morse lattice with central 8-site cluster for (a), $D_4$ symmetric Thue-Morse lattice with central 4-site cluster for (b), $D_2$ symmetric Thue-Morse lattice for (c), $D_5$ symmetric Penrose lattice for (d) and $D_8$ symmetric Ammann-Beenker lattice for (e).
    The solid (hollow) circle denotes the site in sublattice $\mathcal{A}$ ($\mathcal{B}$) and the solid line represents the NN or NNN bond.}
    \label{fig:latt}
\end{figure}

As the beginning of our study, we construct the Hubbard model in several 2D bipartite quasicrystals (QCs).
The geometries of the five QCs we studied are shown in Fig. \ref{fig:latt}, including the $D_4$ symmetric Thue-Morse QC with central 8-site cluster in (a) panel, the $D_4$ symmetric Thue-Morse QC with central 4-site cluster in (b) panel, the $D_2$ symmetric Thue-Morse lattice in (c) panel, the $D_5$ symmetric Penrose QC in (d) panel and the $D_8$ symmetric Ammann-Beenker lattice in (e) panel.

The Hubbard Hamiltonian in the $D_4$ and $D_2$ symmetric Thue-Morse lattices takes the form of
\begin{equation}\label{eqS:H_D4D2}
    H = 
    - t_1 \sum_{\langle i,j\rangle\sigma} \left( c^\dagger_{i\sigma} c_{j\sigma} + \mathrm{h.c.}\right) 
    - t_2 \sum_{\langle\langle i,j\rangle\rangle\sigma} \left( c^\dagger_{i\sigma} c_{j\sigma} + \mathrm{h.c.}\right) 
    + U \sum_{i} n_{i\uparrow} n_{i\downarrow},
\end{equation}
and the one in the $D_5$ symmetric Penrose lattice and $D_8$ symmetric Ammann-Beenker lattice is written as
\begin{equation}\label{eqS:H_D5D8}
    H = 
    - t_1 \sum_{\langle i,j\rangle\sigma} \left( c^\dagger_{i\sigma} c_{j\sigma} + \mathrm{h.c.}\right) 
    + U \sum_{i} n_{i\uparrow} n_{i\downarrow}.
\end{equation}
Here, $\langle i,j\rangle$ and $\langle\langle i,j\rangle\rangle$ denote the nearest neighbor (NN) and the next nearest neighbor (NNN), respectively. 
$c^{(\dagger)}_{i\sigma}$ annihilates (creates) a spin-$\sigma$ electron at site $i$ and $n_{i\sigma}$ denotes the number operator for spin-$\sigma$ electrons at site $i$.
We set $t_1=1$, $t_2=0.5$ and $U=4$ throughout this work.

The 2D Thue-Morse quasicrystals studied in this work are constructed by the direct product of two 1D Thue-Morse chains.
The 1D Thue-Morse sequence $\{t(n)\}$ is defined recursively by
\begin{equation}
\begin{cases}
t(0)=0,\\
t(2n)=t(n),\\
t(2n+1)=1-t(n),\\
\end{cases}
\end{equation}
where $n \geq 0$.

To construct a 2D Thue-Morse structure, we introduce two independent Thue-Morse sequences along the Cartesian axes:
\begin{equation}
t_x(x)=t(x),\qquad t_y(y)=t(y),\qquad x,y\in\mathbb{Z}.
\end{equation}

The direct product of these two sequences defines a bivariate array
\begin{equation}
T:\mathbb{Z}^2\to\{0,1\},\qquad
T(x,y) = \bigl(t(x)+t(y)\bigr)\bmod 2.
\end{equation}

This array $T(x,y)$ exhibits long-range order without full translational periodicity and possesses a self-similar Fourier spectrum, thus serving as a standard direct-product realization of a 2D Thue-Morse quasicrystal. The values $T(x,y)=0$ and $T(x,y)=1$ can be interpreted as two distinct tile types or atomic species arranged on the square lattice.

\begin{figure}[htbp]
    \centering
    \includegraphics[width=0.4\linewidth]{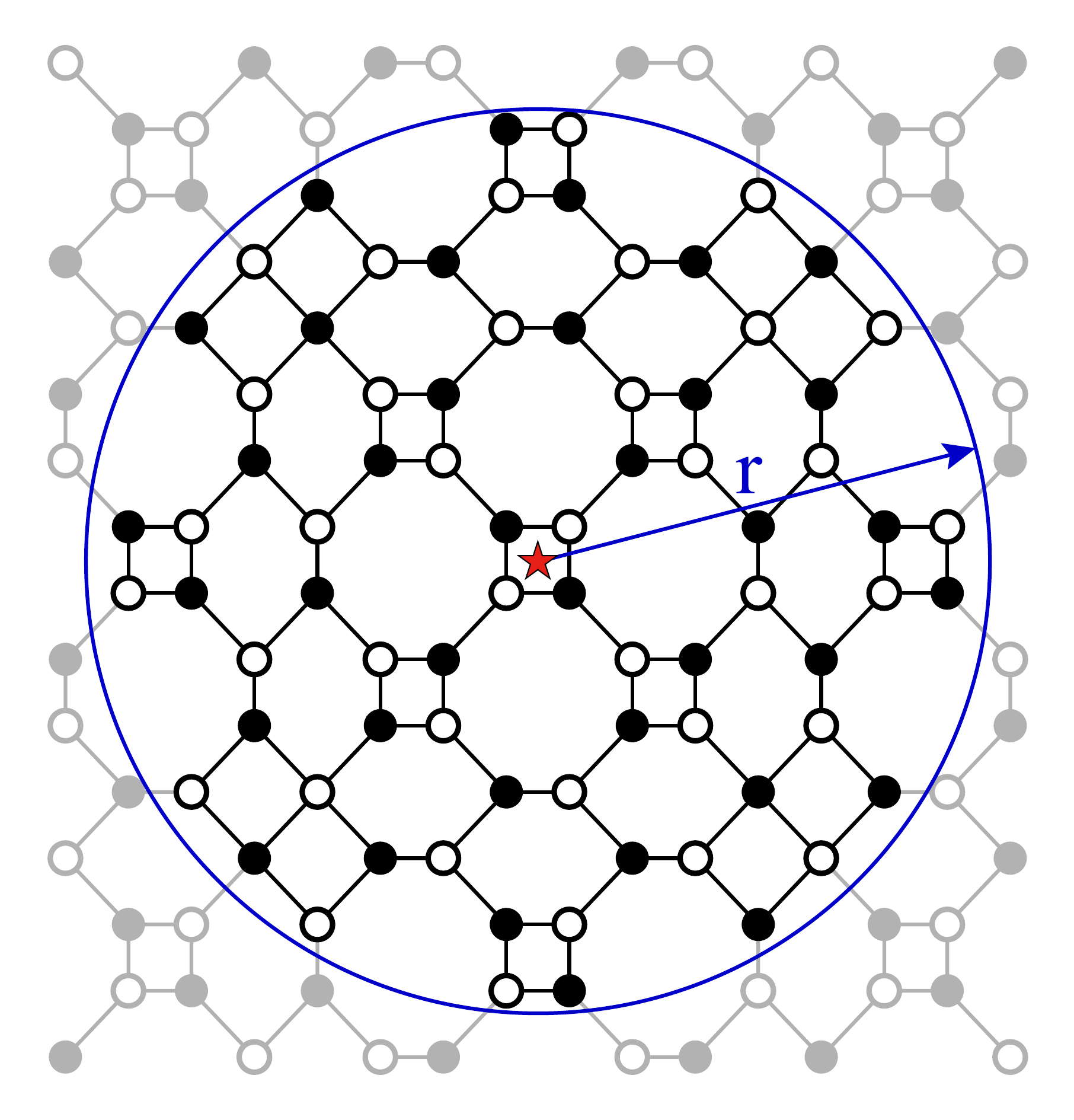}
    \caption{Schematic diagram of 2D Thue-Morse lattice with the radius restriction.
    The semi-transparent sites outside radius $r$ are ignored}
    \label{fig:ThueMorse}
\end{figure}

For convenience and for comparison with the finite Penrose and Ammann-Beenker clusters considered in this work, we restrict the system to a bounded domain centered at the spatial inversion center. Defining the Euclidean norm \(\lVert (x,y)\rVert=\sqrt{x^2+y^2}\), we select only those lattice sites lying within a radius \(r>0\), as shown in Fig. \ref{fig:ThueMorse}. The resulting finite configuration is described by the set
\begin{equation}
\mathcal{T}_r=\bigl\{\, (x,y)\in\mathbb{Z}^2 \;\big|\; T(x,y)\in\{0,1\},\; \lVert (x,y)\rVert\le r \,\bigr\},
\end{equation}
which does not break the point group symmetry or bipartite structure of the Thue-Morse lattice. This radius restriction is only a finite-cluster choice and does not change the underlying direct-product Thue-Morse construction in the thermodynamic limit.
% ----- Sec A [end] -----

% ----- Sec B [begin] -----
\section{B. Projector Quantum Monte Carlo Algorithm with Non-uniform Time Slice Strategy}
\label{appendix_b}

In this section, we propose a non-uniform time slice strategy applicable to the projector quantum Monte Carlo (PQMC) algorithm, which saves more than 40\% of time with the same accuracy compared to the usual uniform time slice strategy.

In PQMC simulations, the expectation of a physical quantity $O$ is written as
\begin{equation}\label{eqS:O}
    \langle O\rangle = \lim_{\Theta\rightarrow\infty} \frac{\langle\psi|\rme^{-\Theta H} O \rme^{-\Theta H}|\psi\rangle}{\langle\psi|\rme^{-2\Theta H}|\psi\rangle},
\end{equation}
where $|\psi\rangle$ represents the trial wave function. 
$\Theta$ denotes the total imaginary time, and $\rme^{-\Theta H}$ is the total imaginary time projection operator. 
The overlap integral between the ground state and the trial wave function $|\langle\Psi_0|\psi\rangle|^2$ can depict the degree of approximation between the two.
$|\langle\Psi_0|\psi\rangle|^2 = 1$ indicates that the trial wave function is the exact ground state, while $|\langle\Psi_0|\psi\rangle|^2 = 0$ implies that the two are orthogonal.
As it gets closer to $0$, the larger $\Theta$ is required to project the trial wave function to reach the ground state.

For systems with many-body interactions, the Hubbard-Stratonovich (HS) decomposition is used to deal with the interaction terms, in which the total projection operator $\rme^{-2\Theta H}$ is decomposed into projection operators with several time slices as
\begin{equation}\label{eqS:Trotter}
    \rme^{-2\Theta H} = \lim_{n\rightarrow\infty} \rme^{-2n\Delta\tau H},
\end{equation}
which is always implemented numerically using the Trotter decomposition and thus introduce the so-called Trotter error.
Obviously, the accuracy of the simulated ground state with the same total imaginary time $\Theta$ is restricted by both the deviation of the trial wave function from the exact ground state and the Trotter error.
When the overlap integral is less than a certain threshold, the deviation of the trial wave function from the exact ground state cannot be adequately eliminated after projection; while the time slices $\Delta\tau$ are too large, the Trotter error leads to insufficient simulation accuracy.
When both are taken into account, the excessive number of time slices limits the computational efficiency of PQMC simulations.

In previous PQMC studies, the time slices $\Delta\tau$ are set to a unitary value, leading to a large number of time slices $\Delta\tau$ for a sufficiently large $\Theta$ and a sufficiently small Trotter error.
Here, we propose a non-uniform time-slice strategy to accelerate the PQMC simulation by setting a series of different $\Delta\tau$.
The strategy decomposes the total projection operator into projection operators with non-uniform time slices as
\begin{align}
    \rme^{-\Theta H} O \rme^{-\Theta H} &= 
    \lim_{n\rightarrow\infty} \rme^{-\Delta\tau_1 H} \rme^{-\Delta\tau_2 H} \cdots \rme^{-\Delta\tau_n H} O \rme^{-\Delta\tau_n H} \rme^{-\Delta\tau_{n-1} H} \cdots \rme^{-\Delta\tau_1 H},\\
    \rme^{-2\Theta H} &= 
    \lim_{n\rightarrow\infty} \rme^{-\Delta\tau_1 H} \rme^{-\Delta\tau_2 H} \cdots \rme^{-\Delta\tau_n H} \rme^{-\Delta\tau_n H} \rme^{-\Delta\tau_{n-1} H} \cdots \rme^{-\Delta\tau_1 H},
\end{align}
where
\begin{equation}
    \Theta = \sum_{i=1}^n \Delta\tau_i
    \quad\text{and}\quad
    \Delta\tau_1\geq\Delta\tau_2\geq\cdots\geq\Delta\tau_n > 0.
\end{equation}

This strategy aims to balance the Trotter error and the error between the trial wave function and the exact ground state by regulating the time slices.
In a many-body system, particularly a strongly correlated system (e.g., large-$U$ Hubbard model), the trial wave function $|\psi\rangle$ always deviates far from the exact ground state $|\Psi_0\rangle$, which requires a large $\Theta$ to cool $|\psi\rangle$ to $|\Psi_0\rangle$.
Thus, it is a sensible strategy to preferentially apply the projection operator with larger time slices to the trial wave function for a rapid approach to the ground state, and then apply the ones with smaller time slices to reduce the Trotter error and further optimize the state.

\begin{figure}[htbp]
    \centering
    \includegraphics[width=0.8\linewidth]{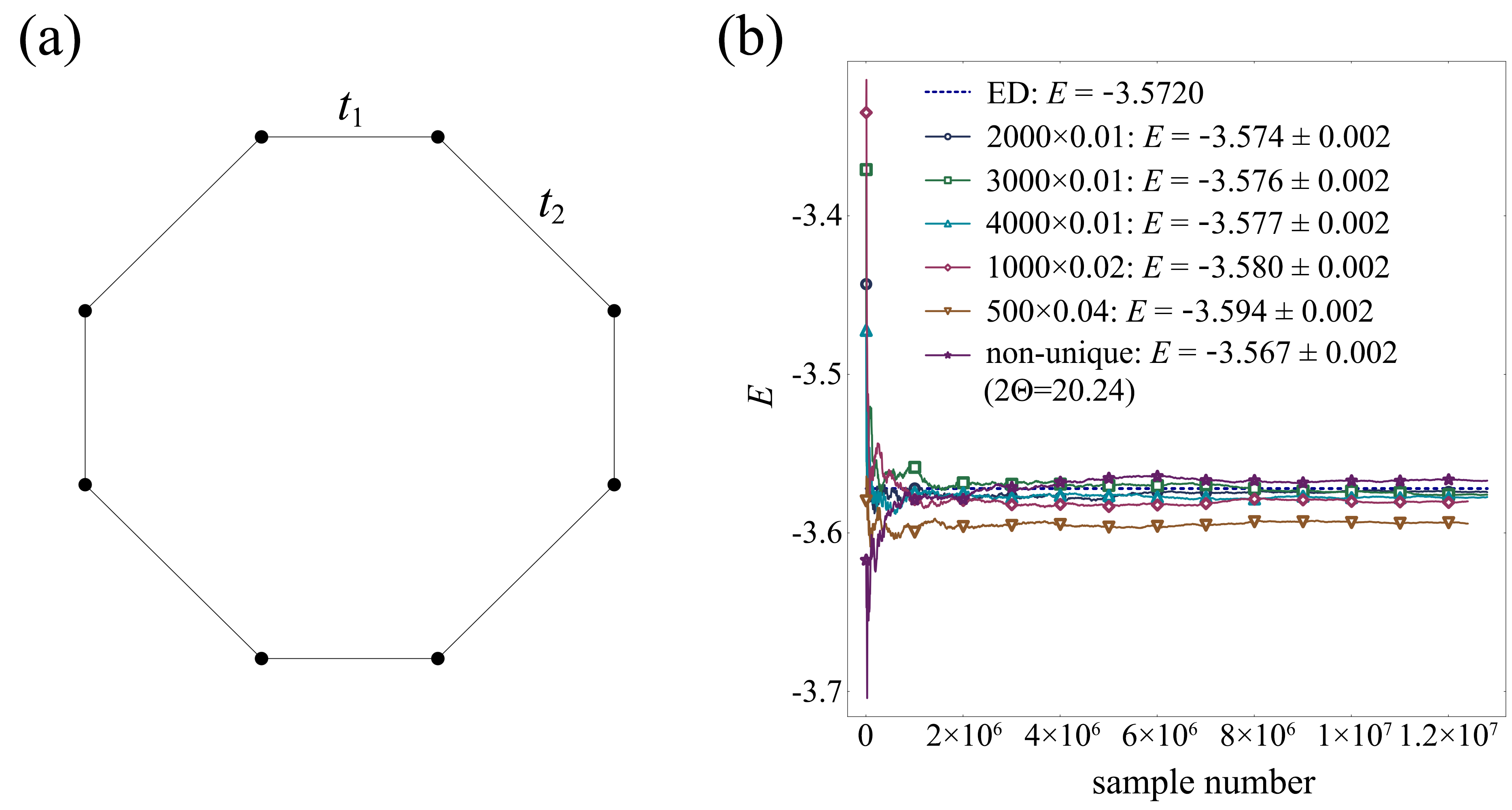}
    \caption{
    (a) The schematic diagram of 8-site Thue-Morse lattice. 
    The solid line represents the NN or NNN bond.
    (b) The energy as a function of sampling number.
    The exact ground state energy and the statistical mean and error of the energy for different parameters are labeled in the figure.}
    \label{fig:benchmark}
\end{figure}

In order to demonstrate the efficiency of the non-uniform strategy, we do a series of tests with the Hubbard model Eq. (\ref{eqS:H_D4D2}) in the 8-site Thue-Morse lattice, while comparing with the results of exact diagonalization (ED). 
Fig. \ref{fig:benchmark} (a) illustrates the geometry of the 8-site Thue-Morse lattice, where the solid lines indicate the NN or NNN bonds.

We solve the ground state for $2\Theta\approx20$ with $600$ non-uniform time slices, i.e., $2\Theta = 2 \times (20\times0.2 + 30\times0.1 + 30\times0.05 + 30\times0.02 + 50\times0.01 + 80\times0.005 + 60\times0.002) = 20.24$, compared to the results for $2\Theta = 20$ with unitary time slices $\Delta\tau = 0.01$, $0.02$ and $0.04$ and for $2\Theta = 30$ and $40$ with unitary time slice $\Delta\tau=0.01$.

The results are shown in Fig. \ref{fig:benchmark} (b).
Comparison of results for $2\Theta=20$, $30$ and $40$ with unitary time slices $\Delta\tau = 0.01$ indicates that $2\Theta=20$ is sufficient to allow the trial wave function to evolve to the ground state, while the accumulated Trotter error amplifies the systematic error for $2\Theta = 30$ and $40$.
Combining the results for $2\Theta=20$ with uniform $\Delta\tau=0.01$, $0.02$ and $0.04$, it is easy to see that the increase in $\Delta\tau$ leads to a rapid deviation of the energy from the exact ground state due to the increasing Trotter error.
The results of the non-uniform time slice strategy with $600$ time slices demonstrate similar accuracy to the results for $2\Theta=20$ with uniform $\Delta\tau=0.02$, slightly worse than the one with $\Delta\tau=0.01$.
As a result, our non-uniform time slice strategy saves more than 40\% of time with the same accuracy compared to the usual uniform time slice strategy, exhibiting a great value for numerate simulation.

In fact, we do not extensively test the performance of different sets of non-uniform time slices.
We believe that more reasonable parameter selection allows this strategy to at least double the performance of the PQMC algorithm, particularly in a strongly correlated system.
% ----- Sec B [end] -----

% ----- Sec C [begin] -----
\section{C. Trial wave function in the PQMC simulation}
\label{appendix_c}

The trial wavefunction is introduced as the initial state for imaginary time propagation in PQMC simulations.
For the half-filled Hubbard model, it is a common and convenient construction of the trial wave function as follows:
We denote the tight-binding (TB) Hamiltonian (including hopping terms and pinning field terms, see Sec. \textbf{E} and \textbf{F}) as $H_\text{TB}$, which can be decomposed to spin-$\uparrow$ ($H_\text{TB}^\uparrow$) and spin-$\downarrow$ ($H_\text{TB}^\downarrow$) components as $H_\text{TB} = H_\text{TB}^\uparrow + H_\text{TB}^\downarrow$.
The coefficient matrix of $H_\text{TB}$ is a block-diagonal matrix with $H_\text{TB}^\uparrow$ and $H_\text{TB}^\downarrow$ as diagonal blocks, i.e.,
\begin{equation}\label{eqS:HK}
[H_\text{TB}] = \left[
\begin{array}{c c}
    [H_\text{TB}^\uparrow] &  \\
     & [H_\text{TB}^\downarrow]
\end{array}\right].
\end{equation}
Here, $[\dots]$ denotes the coefficient matrix. 
Then we solve the eigen equations
\begin{equation}\label{eqS:eigen_up}
[H_\text{TB}^\uparrow] [\phi^\uparrow] \equiv
[H_\text{TB}^\uparrow] \left[\begin{array}{c c c c}
    \phi_{1,1}^\uparrow & \phi_{1,2}^\uparrow & \cdots & \phi_{1,n_\uparrow}^\uparrow \\
    \phi_{2,1}^\uparrow & \phi_{2,2}^\uparrow & \cdots & \phi_{2,n_\uparrow}^\uparrow \\ 
    \vdots & \vdots & \ddots & \vdots \\
    \phi_{n,1}^\uparrow & \phi_{n,2}^\uparrow & \cdots & \phi_{n,n_\uparrow}^\uparrow
\end{array}\right]
= \left[\begin{array}{c c c c}
    \phi_{1,1}^\uparrow & \phi_{1,2}^\uparrow & \cdots & \phi_{1,n_\uparrow}^\uparrow \\
    \phi_{2,1}^\uparrow & \phi_{2,2}^\uparrow & \cdots & \phi_{2,n_\uparrow}^\uparrow \\ 
    \vdots & \vdots & \ddots & \vdots \\
    \phi_{n,1}^\uparrow & \phi_{n,2}^\uparrow & \cdots & \phi_{n,n_\uparrow}^\uparrow
\end{array}\right]
\left[\begin{array}{c c c c}
    E_1^\uparrow & & & \\
     & E_2^\uparrow & & \\ 
     &  & \ddots & \\
     &  &  & E_{n_\uparrow}^\uparrow
\end{array}\right]
\end{equation}
and
\begin{equation}\label{eqS:eigen_dn}
[H_\text{TB}^\downarrow] [\phi^\downarrow] \equiv
[H_\text{TB}^\downarrow] \left[\begin{array}{c c c c}
    \phi_{1,1}^\downarrow & \phi_{1,2}^\downarrow & \cdots & \phi_{1,n_\downarrow}^\downarrow \\
    \phi_{2,1}^\downarrow & \phi_{2,2}^\downarrow & \cdots & \phi_{2,n_\downarrow}^\downarrow \\ 
    \vdots & \vdots & \ddots & \vdots \\
    \phi_{n,1}^\downarrow & \phi_{n,2}^\downarrow & \cdots & \phi_{n,n_\downarrow}^\downarrow
\end{array}\right]
= \left[\begin{array}{c c c c}
    \phi_{1,1}^\downarrow & \phi_{1,2}^\downarrow & \cdots & \phi_{1,n_\downarrow}^\downarrow \\
    \phi_{2,1}^\downarrow & \phi_{2,2}^\downarrow & \cdots & \phi_{2,n_\downarrow}^\downarrow \\ 
    \vdots & \vdots & \ddots & \vdots \\
    \phi_{n,1}^\downarrow & \phi_{n,2}^\downarrow & \cdots & \phi_{n,n_\downarrow}^\downarrow
\end{array}\right]
\left[\begin{array}{c c c c}
    E_1^\downarrow & & & \\
     & E_2^\downarrow & & \\ 
     &  & \ddots & \\
     &  &  & E_{n_\downarrow}^\downarrow
\end{array}\right],
\end{equation}
where $n$ denotes the number of lattice sites and $n_\uparrow = \frac{n}{2}$ ($n_\downarrow = \frac{n}{2}$) denotes the particle number of spin-$\uparrow$ (spin-$\downarrow$) Fermion. 
$E_i^\uparrow$ ($E_i^\downarrow$) is the $i$-th lowest eigenvalue of $H_\text{TB}^\uparrow$ ($H_\text{TB}^\downarrow$) with the eigenvector $\phi_i^+$ ($\phi_i^-$) whose coefficient matrix is the $i$-th column of $[\phi^\uparrow]$ ($[\phi^\downarrow]$).

We construct the trial wave function $|\psi\rangle$ as
\begin{equation}\label{eqS:psi_zero}
    |\psi\rangle = \prod_{m=1}^{N/2}\prod_{\sigma} \sum_{i=1}^N \phi_{im}^{\sigma} c^\dagger_{i\sigma} |0\rangle,
\end{equation}
with the coefficient matrix
\begin{equation}\label{eqS:psi_zero_c}
    [\psi] = \left[
    \begin{array}{c c}
        [\phi^\uparrow] &  \\
         & [\phi^\downarrow]
    \end{array}
    \right] = \left[
    \begin{array}{c c c c c c c c}
        \phi_{1,1}^\uparrow & \phi_{1,2}^\uparrow & \cdots & \phi_{1,n_\uparrow}^\uparrow & & & & \\
        \phi_{2,1}^\uparrow & \phi_{2,2}^\uparrow & \cdots & \phi_{2,n_\uparrow}^\uparrow & & & & \\ 
        \vdots & \vdots & \ddots & \vdots & & & & \\
        \phi_{n,1}^\uparrow & \phi_{n,2}^\uparrow & \cdots & \phi_{n,n_\uparrow}^\uparrow & & & & \\
        & & & & \phi_{1,1}^\downarrow & \phi_{1,2}^\downarrow & \cdots & \phi_{1,n_\downarrow}^\downarrow \\
        & & & & \phi_{2,1}^\downarrow & \phi_{2,2}^\downarrow & \cdots & \phi_{2,n_\downarrow}^\downarrow \\ 
        & & & & \vdots & \vdots & \ddots & \vdots \\
        & & & & \phi_{n,1}^\downarrow & \phi_{n,2}^\downarrow & \cdots & \phi_{n,n_\downarrow}^\downarrow
    \end{array}
    \right].
\end{equation}

The construction is sign-problem-free in the lattice with a gapped Fermi surface (e.g. $D_2$ and $D_4$ symmetric Thue-Morse lattice).
However, the sign-problem occurs in the half-filled Hubbard model with a gapless ground state of $H_\text{TB}$ (e.g. $D_5$ symmetric Penrose lattice or $D_8$ symmetric Ammann-Beenker lattice) because of the incompleteness of the trial wave function (\ref{eqS:psi_zero}), which can be avoided by the particle-hole (PH) transformation.

Another character of the trial wave function (\ref{eqS:psi_zero}) is the fixed and equal particle numbers of spin-$\uparrow$ and spin-$\downarrow$, which requires that the net magnetism of the measured ground state be zero.
Without the magnetic pinning field, the spin $\mathrm{SU(2)}$ symmetry preserves the measured ground state with zero net magnetism in finite size.
However, the introduction of magnetic pinning field explicitly breaks spin $\mathrm{SU(2)}$ and $\mathcal{T}$ symmetries, leading to a possible non-zero net magnetism and the particle number difference $\langle n_\uparrow\rangle - \langle n_\downarrow\rangle$ that cannot be determined a priori.
Therefore, we propose a new and complete construction to capture different N\'eel orders, which avoids the sign-problem without the aid of PH transformation.

We denote the TB Hamiltonian without pinning field as $H_{\text{TB}0} = H_{\text{TB}0}^\uparrow + H_{\text{TB}0}^\downarrow$, where $H_{K0}^\uparrow$ ($H_{K0}^\downarrow$) is the spin-$\uparrow$ (spin-$\downarrow$) component.
The coefficient matrix follows
\begin{equation}\label{HK0}
    [H_{\text{TB}0}] = \left[
\begin{array}{c c}
    [H_{\text{TB}0}^\uparrow] &  \\
     & [H_{\text{TB}0}^\downarrow]
\end{array}\right].
\end{equation}
Here, $[H_{\text{TB}0}^\uparrow] = [H_{\text{TB}0}^\downarrow]$.
Different from Eqs. (\ref{eqS:eigen_up}) and (\ref{eqS:eigen_dn}), the eigen equations are solved by
\begin{equation}\label{eqS:eigen_up0}
[H_{\text{TB}0}^\sigma] [\phi] \equiv
[H_{\text{TB}0}^\sigma] \left[\begin{array}{c c c c}
    \phi_{1,1} & \phi_{1,2} & \cdots & \phi_{1,n} \\
    \phi_{2,1} & \phi_{2,2} & \cdots & \phi_{2,n} \\ 
    \vdots & \vdots & \ddots & \vdots \\
    \phi_{n,1} & \phi_{n,2} & \cdots & \phi_{n,n}
\end{array}\right]
= \left[\begin{array}{c c c c}
    \phi_{1,1} & \phi_{1,2} & \cdots & \phi_{1,n} \\
    \phi_{2,1} & \phi_{2,2} & \cdots & \phi_{2,n} \\ 
    \vdots & \vdots & \ddots & \vdots \\
    \phi_{n,1} & \phi_{n,2} & \cdots & \phi_{n,n}
\end{array}\right]
\left[\begin{array}{c c c c}
    E_1 & & & \\
     & E_2 & & \\ 
     &  & \ddots & \\
     &  &  & E_{n}
\end{array}\right].
\end{equation}
The trial wave function $|\psi\rangle$ are constructed by
\begin{equation}\label{eqS:psi_nonzero}
    |\psi\rangle = \prod_{m=1}^{N} \sum_{i=1}^N \phi_{im} \left(c^\dagger_{i\uparrow} + c^\dagger_{i\downarrow}\right) |0\rangle,
\end{equation}
with the coefficient matrix
\begin{equation}\label{eqS:psi_nonzero_c}
    [\psi] = \left[
\begin{array}{c c c c c c c c}
    \phi_{1,1} & \phi_{1,2} & \cdots & \phi_{1,n} \\
    \phi_{2,1} & \phi_{2,2} & \cdots & \phi_{2,n} \\ 
    \vdots & \vdots & \ddots & \vdots\\
    \phi_{n,1} & \phi_{n,2} & \cdots & \phi_{n,n} \\
    \phi_{1,1} & \phi_{1,2} & \cdots & \phi_{1,n} \\
    \phi_{2,1} & \phi_{2,2} & \cdots & \phi_{2,n} \\ 
    \vdots & \vdots & \ddots & \vdots \\
    \phi_{n,1} & \phi_{n,2} & \cdots & \phi_{n,n}
\end{array}
\right].
\end{equation}

In our subsequent study, we confirm the existence of altermagnetism (AM) and antiferromagnetism (AFM) N\'eel order with a zero net magnetism in $D_4$ and $D_2$ symmetric Thue-Morse lattice using the trial wave function (\ref{eqS:psi_nonzero}).
As a result, we reconsider the trial wave function (\ref{eqS:psi_zero}) to accelerate the simulations.
% ----- Sec C [end] -----

% ----- Sec D [begin] -----
\section{D. Spin-Resolved Angle-Resolved Photoemission Spectrum}
\label{appendix_d}

To distinguish the AM and AFM states, we consider the spin-resolved angle-resolved photoemission spectrum\cite{chen2025quasi,li2025quasi,dornellas2025alter}. 

We begin with a mean-field Hamiltonian with magnetic order, 
\begin{equation}
    H_{\mathrm{MF}} = 
    - t_1 \sum_{\langle i,j\rangle\sigma} \left( c^\dagger_{i\sigma} c_{j\sigma} + \mathrm{h.c.}\right) 
    - t_2 \sum_{\langle\langle i,j\rangle\rangle\sigma} \left( c^\dagger_{i\sigma} c_{j\sigma} + \mathrm{h.c.}\right) 
    - \mu \sum_{i\sigma} n_{i\sigma} 
    - \frac{2U}{3} \sum_{i} \left\langle S_i^z \right\rangle \left(n_{i\uparrow}-n_{i\downarrow}\right). 
\end{equation}
For Thue-Morse lattice, $t_1=1$ and $t_2=0.5$. 
In the Wannier basis, the matrix form of $H_{\mathrm{MF}}$ is block-diagonal, 
\begin{equation}
    \left[H_{\mathrm{MF}}\right] = 
    \left[
    \begin{array}{c c}
        [H_{\uparrow}] &  \\
         & [H_{\downarrow}]
    \end{array}
    \right], 
\end{equation}
where $\left[H_{\mathrm{MF}}\right]$ is the matrix form of $H_{\mathrm{MF}}$ in the Wannier basis, and $\left[H_{\sigma}\right]$ is the matrix form of the mean-field Hamiltonian for spin $\sigma$, which is defined as 
\begin{equation}
    H_{\sigma} = 
    - t_1 \sum_{\langle i,j\rangle} \left( c^\dagger_{i\sigma} c_{j\sigma} + \mathrm{h.c.}\right) 
    - t_2 \sum_{\langle\langle i,j\rangle\rangle} \left( c^\dagger_{i\sigma} c_{j\sigma} + \mathrm{h.c.}\right) 
    -\mu \sum_{i} n_{i\sigma}
    - \frac{2U}{3} \sigma \sum_{i} \left\langle S_i^z \right\rangle n_{i\sigma}. 
\end{equation}

The spectrum weight for spin $\sigma$ at a certain energy level $\omega$ is expressed as
\begin{equation}
    \mathcal{A}_{\sigma}(\bm{p},\omega) = -\frac{\eta}{\pi} \frac{\left|\left\langle{}\bm{p}\sigma|m\sigma\right\rangle\right|^2}{\left(\omega-E_m^{\sigma}\right)^2+\eta^2}, 
\end{equation}
where $E_m^{\sigma}$ and $\left|m\sigma\right\rangle$ are the $m$-th eigenvalue and the corresponding eigenstate of $H_{\sigma}$, $\left|\bm{p}\sigma\right\rangle=\frac{1}{\sqrt{N}}\sum_{i}\mathrm{e}^{\mathrm{i}\bm{p}\cdot\bm{r}_{i}}c^{\dagger}_{i\sigma}\left|0\right\rangle$ ($N$ is the number of sites, $\bm{r}_i$ is the position of site $i$ and $\left|0\right\rangle$ is vacuum state) is the plane wave with momentum $\bm{p}$, and $\eta$ is a small Lorentzian expansion width. 
% ----- Sec D [end] -----

% ----- Sec E [begin] -----
\section{E. More Details and Results with Local Magnetic Pinning Field}
\label{appendix_e}

The PQMC algorithm is limited by the inability to distinguish between degenerate ground states, and therefore cannot correctly compute the nonzero order parameters for spontaneous symmetry breaking. 
In addition, the correlation function cannot accurately evaluate the magnitude of the local order parameters for further spectroscopic study.
Therefore, we introduce the pinning field and obtain the exact magnitude and distribution of the order parameters by extrapolation within the linear response region.

In this section, we apply the local magnetic pinning field of different patterns on the central small-cluster or single-site to determine the pattern of the ground state. 
The local magnetic pinning field is written as
\begin{equation}\label{eqS:Hpl}
    H_p^\text{local} = h_l \sum_{i\in\mathcal{C}} {\text{sign}(i)}\ S^z_{i},
\end{equation}
where $\mathcal{C}$ represents the central small-cluster or single-site and $\text{sign}(i)$ denotes the sign of the pinning field at the site $i$. $S^z_i$ is the $z$-component spin operator at site $i$. 
The total Hamiltonian is written as
\begin{equation}\label{eqS:Htot}
    H = 
    - t_1 \sum_{\langle i,j\rangle\sigma} \left( c^\dagger_{i\sigma} c_{j\sigma} + \mathrm{h.c.}\right) 
    - t_2 \sum_{\langle\langle i,j\rangle\rangle\sigma} \left( c^\dagger_{i\sigma} c_{j\sigma} + \mathrm{h.c.}\right) 
    + U \sum_{i} n_{i\uparrow} n_{i\downarrow}
    + H_p^\text{local}.
\end{equation}

\begin{figure}[htbp]
    \centering
    \includegraphics[width=0.98\linewidth]{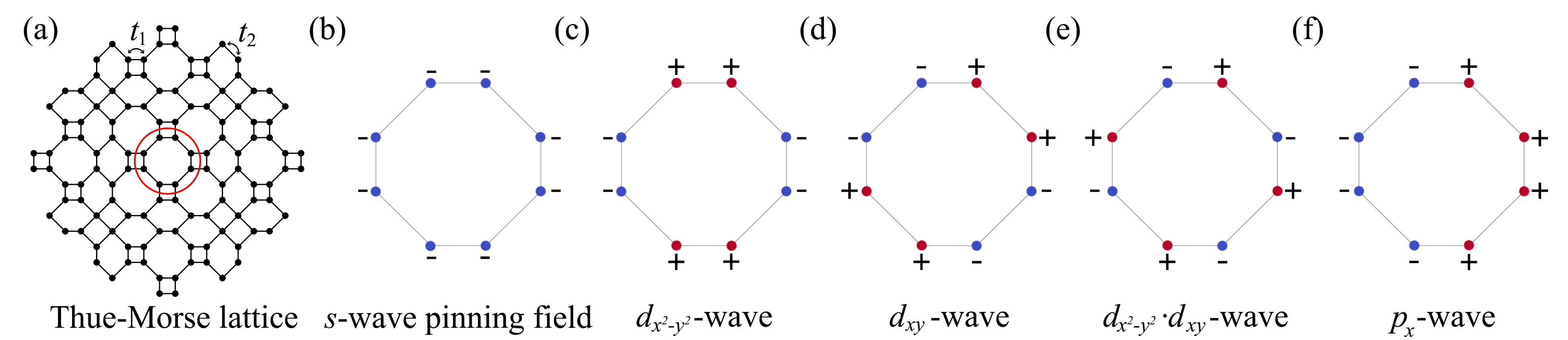}
    \caption{
    (a) The schematic diagram of 120-site $D_4$ symmetric Thue-Morse lattice with central 8-site cluster circled in red. 
    The black dot denotes the site and the solid line represents the NN or NNN bond. 
    The local magnetic pinning field is applied in the cluster circled in red. 
    (b-f) The schematic diagram of the local magnetic pinning field signs of $s$-wave for (b), $d_{x^2-y^2}$-wave for (c), $d_{xy}$-wave for (d), $d_{x^2-y^2}\cdot d_{xy}$-wave for (e), $p_x$-wave pattern for (f).}
    \label{fig:local_D4with8sites_latt}
\end{figure}

Here, we take the $D_4$ symmetric Thue-Morse lattice with central 8-site cluster as an example. 
Fig. \ref{fig:local_D4with8sites_latt} (a) exhibits the structure of 120-site $D_4$ symmetric Thue-Morse lattice with central 8-site cluster circled in red, where the black dot represents the site and the solid line represents the NN or NNN bond. 
Figs. \ref{fig:local_D4with8sites_latt} (b-f) show the signs of the local magnetic pinning field with $s$-wave, $d_{x^2-y^2}$-wave, $d_{xy}$-wave, $d_{x^2-y^2}\cdot d_{xy}$-wave, $p_x$-wave pattern, respectively. 
We set $t_1=1$, $t_2=0.5$, $U=4$ and $h_l=0.2$ to determine the ground state pattern of the original Hamiltonian (\ref{eqS:H_D4D2}) without the pinning field.

\begin{figure}[htbp]
    \centering
    \includegraphics[width=0.98\linewidth]{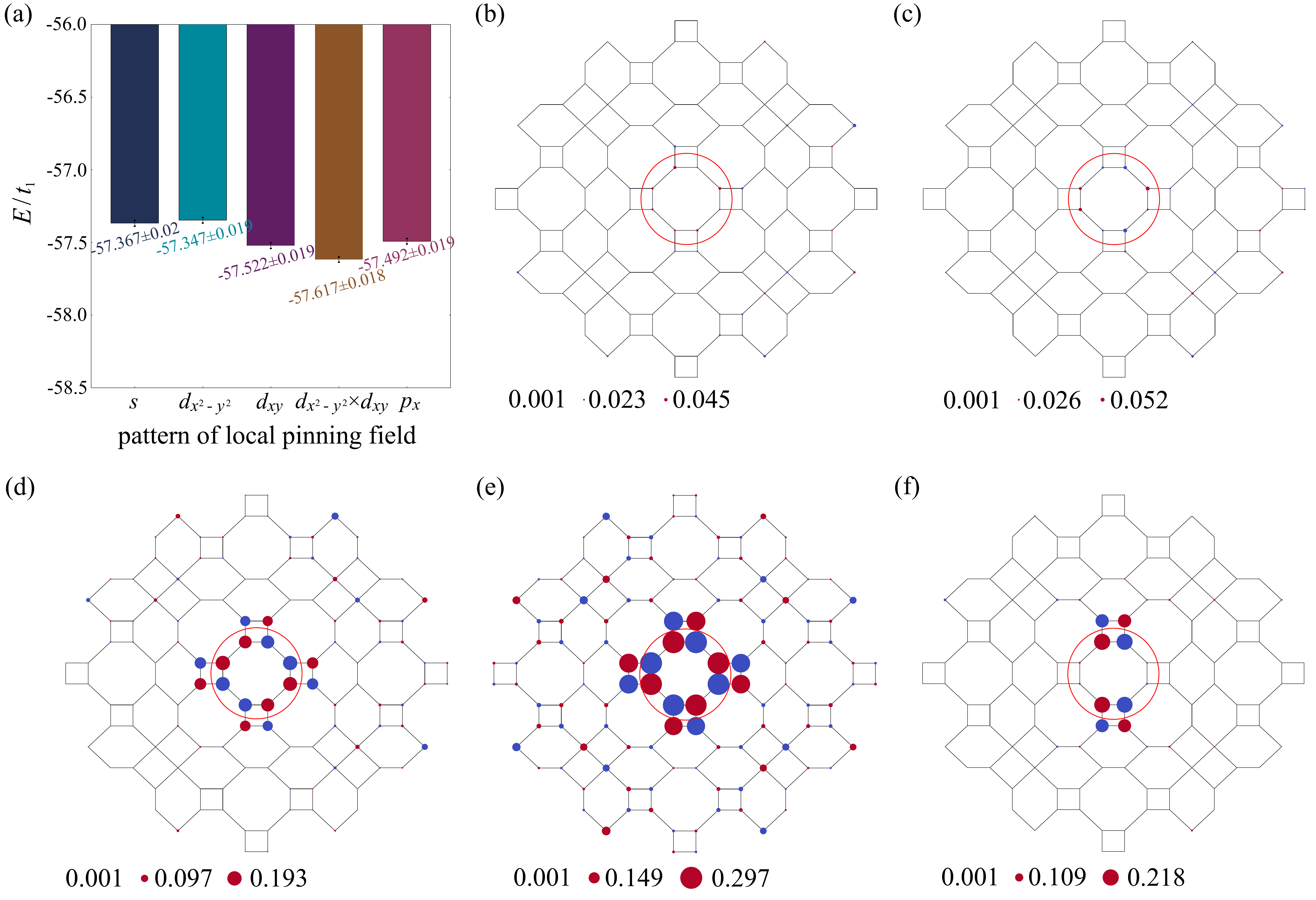}
    \caption{
    (a) The ground state energies in the $D_4$ symmetric Thue-Morse lattice under local magnetic pinning fields with different patterns on the central 8-site cluster (circled in red). 
    (b-f) The distribution of the magnetic order parameter $\langle S^z_i\rangle$ under local magnetic pinning fields with $s$-wave for (b), $d_{x^2-y^2}$-wave for (c), $d_{xy}$-wave for (d), $d_{x^2-y^2}\cdot d_{xy}$-wave for (e), $p_x$-wave pattern for (f)}
    \label{fig:local_D4with8sites}
\end{figure}

Fig. \ref{fig:local_D4with8sites} shows the results under local pinning field with different patterns. 
As shown in Fig. \ref{fig:local_D4with8sites} (a), the ground state of the Hamiltonian (\ref{eqS:Htot}) with the local magnetic pinning field with $d_{x^2-y^2}\cdot d_{xy}$ pattern holds the lowest energy. 
As the introduction of a weak local pinning field lifts the energy of the ground state mismatching the pattern of the pinning field and lowers the matched one, the results indicate that the ground state of the original Hamiltonian (\ref{eqS:H_D4D2}) without the pinning field takes $d_{x^2-y^2}\cdot d_{xy}$ pattern.

Figs. \ref{fig:local_D4with8sites} (b-f) exhibit the distribution of the magnetic order parameter $\langle S^z_i\rangle$ under local magnetic pinning fields with $s$-wave, $d_{x^2-y^2}$-wave, $d_{xy}$-wave, $d_{x^2-y^2}\cdot d_{xy}$-wave, $p_x$-wave pattern, respectively. 
The results indicate that the order parameter $\langle S^z_i\rangle$ under the pinning field with $d_{x^2-y^2}\cdot d_{xy}$ pattern maintains a finite value within the finite size, while the ones under the pinning fields with other patterns quickly fall to zero with the increasing distance from the central cluster. 
Since the effect of a local pinning field decays algebraically with distance\cite{assaad2013localpin,parisen2017localpin}, a weak mismatched local pinning field can only pin the local order parameter, while the matched one can stabilize the global order parameter by the existence of the long-range order.
The behavior of the distribution of the magnetic order parameter $\langle S^z_i\rangle$ in finite size also suggests that the ground state of the original Hamiltonian (\ref{eqS:H_D4D2}) without the pinning field holds the $d_{x^2-y^2}\cdot d_{xy}$-wave pattern.

In short, both the lowest energy and the finite value of $\langle S^z_i\rangle$ in finite size indicate that $d_{x^2-y^2}\cdot d_{xy}$-wave is the ground state pattern of the original Hamiltonian (\ref{eqS:H_D4D2}) without the pinning field.

\begin{figure}[htbp]
    \centering
    \includegraphics[width=0.98\linewidth]{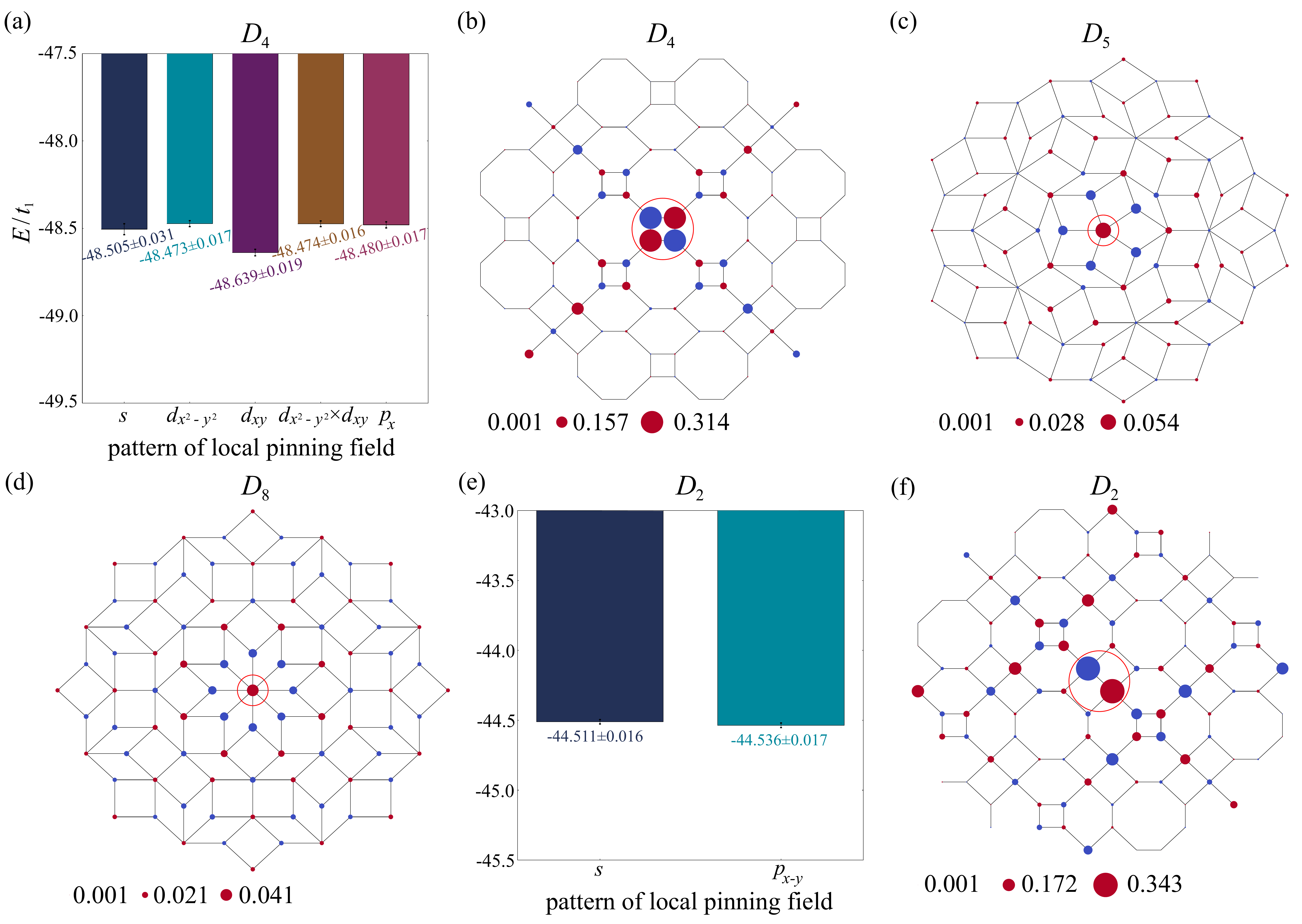}
    \caption{
    (a) The ground state energies in the $D_4$ symmetric Thue-Morse lattice under local magnetic pinning fields with different patterns on the central 4-site cluster. 
    (b) The distribution of the magnetic order parameter $\langle S^z_i\rangle$ under local magnetic pinning field with $d_{xy}$-wave (circled in red) in the $D_4$ symmetric lattice in the same lattice as in (a). 
    (c) The distribution of $\langle S^z_i\rangle$ under local pinning field at the central single-site (circled in red) in the $D_5$ symmetric lattice. 
    (d) The distribution of $\langle S^z_i\rangle$ under local pinning field (circled in red) at the central single-site in the $D_8$ symmetric lattice. 
    (e) The ground state energies in the $D_2$ symmetric Thue-Morse lattice under local magnetic pinning fields with different patterns on the central 2-site cluster. 
    (f) The distribution of the magnetic order parameter $\langle S^z_i\rangle$ under local magnetic pinning field with $p_{x-y}$-wave (circled in red) in the $D_2$ symmetric lattice in the same lattice as in (e)}
    \label{fig:local_other}
\end{figure}

Similarly, we investigate the ground state properties of different QCs under the corresponding local pinning fields and determine their ground state patterns.
Some of the results are shown in Fig. \ref{fig:local_other}. 
Fig. \ref{fig:local_other} (a) exhibits that $d_{xy}$-wave is the pattern of local pinning field with the lowest ground state energy in the $D_4$ symmetric Thue-Morse lattice with central 4-site cluster, which also leads to a finite value of $\langle S^z_i\rangle$ in the finite size in Fig. \ref{fig:local_other} (b) in the same lattice, indicating that $d_{xy}$-wave is the pattern of the ground state of the original Hamiltonian (\ref{eqS:H_D4D2}) without the pinning field. 
Fig. \ref{fig:local_other} (c) indicates that the distribution of $\langle S^z_i\rangle$ in the $D_5$ symmetric Penrose lattice maintains global symmetry breaking with the $s$-wave pattern, suggesting that the ground state pattern is $s$-wave,
while Fig. \ref{fig:local_other} (d) implies the pattern in the $D_8$ symmetric lattice is $s$-wave as well.
Figs. \ref{fig:local_other} (e-f) provide similar information in the $D_2$ symmetric Thue-Morse lattice, which demonstrates the ground state pattern is $p_{x-y}$-wave.
% ----- Sec E [end] -----

% ----- Sec F [begin] -----
\section{F. More Details and Results with Global Magnetic Pinning Field}
\label{appendix_f}

At the end of Sec. \textbf{E}, we have obtained the ground state patterns in different QCs by the application of the local pinning field. 
Although the matched local pinning field globally maintains nonzero order parameters, the convergence rate near the lattice edge is unsatisfactory with significant statistical error.

In this section, we introduce the global pinning fields with the same patterns as the ground states, which demonstrate better global convergence than the local one.
The global pinning field is written as
\begin{equation}\label{eqS:Hpg}
    H_p^\text{global} = h_g \sum_{i\in\mathcal{L}} {\text{sign}(i)}\ S^z_{i},
\end{equation}
where $\mathcal{L}$ denotes the entire lattice. 
As all patterns determined in the different QCs take the form of the N\'eel order, we actually add the global pinning field with the same form in these bipartite lattices, i.e., one sublattice satisfies $\text{sign}(i)=1$ and the other satisfies $\text{sign}(i)=-1$.

\begin{figure}[htbp]
    \centering
    \includegraphics[width=0.76\linewidth]{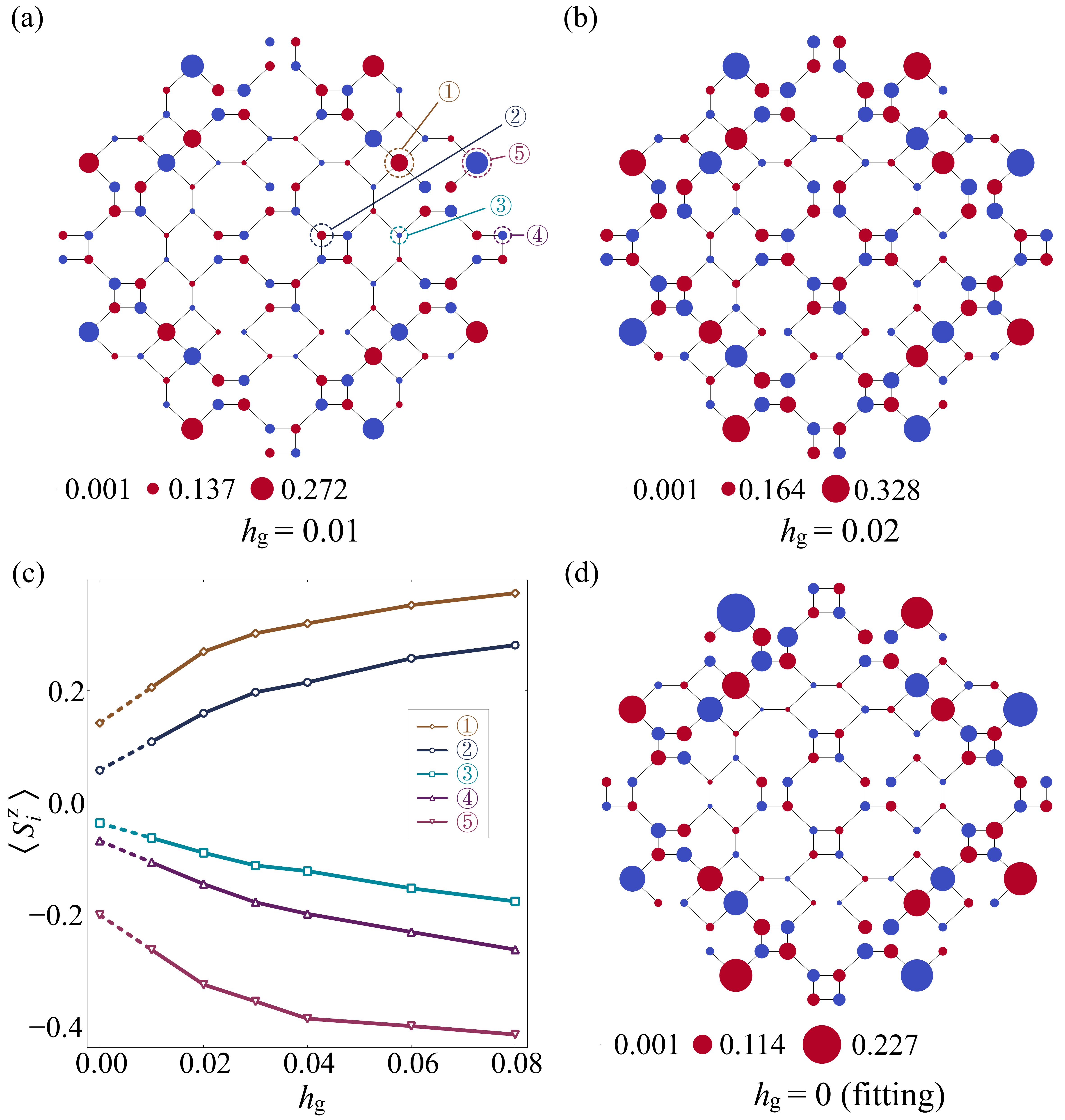}
    \caption{
    (a-b) The distribution of the magnetic order parameter $\langle S^z_i\rangle$ under global magnetic pinning field $h_g = 0.01$ for (a) and $h_g = 0.02$ for (b). 
    (c) The distribution of $\langle S^z_i\rangle$ as a function of the pinning field $h_g$ at the sites $i$ circled in (a). 
    (d) The distribution of $\langle S^z_i\rangle$ fitted by the linear-fitting method.
    The values of $\langle S^z_i\rangle$ are labeled in the figure.}
    \label{fig:global_D4with8sites}
\end{figure}

\begin{figure}[htbp]
    \centering
    \includegraphics[width=0.98\linewidth]{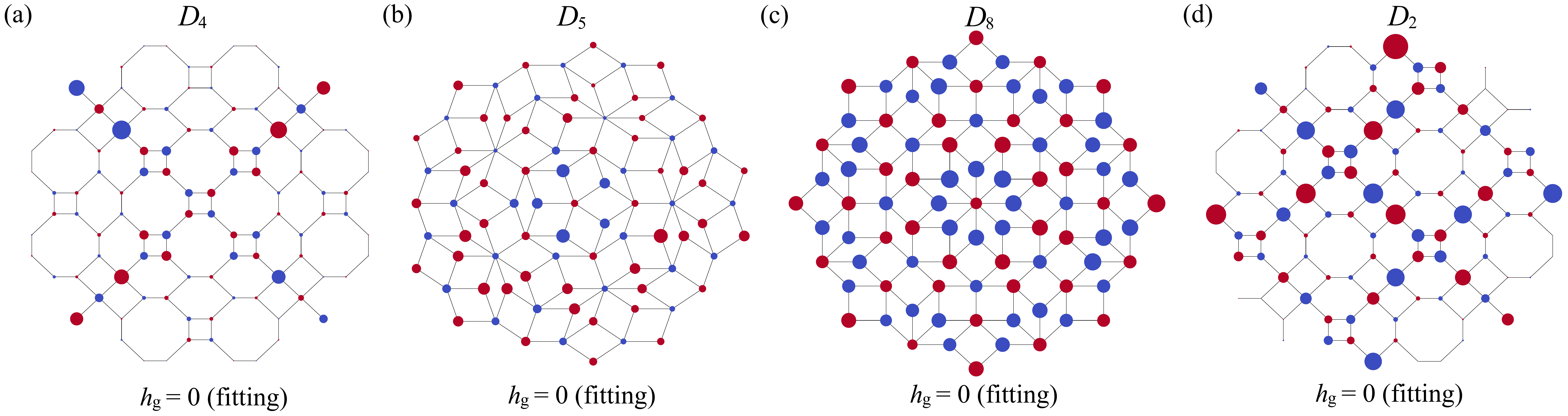}
    \caption{
    (a-d) The distribution of $\langle S^z_i\rangle$ fitted by the linear-fitting method in the $D_4$ symmetric Thue-Morse lattice with central 4-site cluster for (a), $D_5$ symmetric lattice for (b), $D_8$ symmetry for (c) and $D_2$ symmetric Thue-Morse lattice for (d).
    The values of $\langle S^z_i\rangle$ are labeled in the figure.}
    \label{fig:global_other}
\end{figure}

Here, we still take the $D_4$ symmetric Thue-Morse lattice with central 8-site cluster as an example, with the typical parameter set of $t_1=1$, $t_2=0.5$ and $U=4$.

We calculate the magnetic order parameter $\langle S^z_i\rangle$ under the global pinning fields of different strengths $h_g = 0.01$, $0.02$, $0.03$, $0.04$, $0.06$ and $0.08$. The results of $h_g = 0.01$ and $0.02$ are shown in Figs. \ref{fig:global_D4with8sites} (a) and (b), respectively, both of which show excellent global convergence with acceptable statistical error. 
Fig. \ref{fig:global_D4with8sites} (c) illustrates the magnetic order parameter $\langle S^z_i\rangle$ as a function of the pinning field $h_g$ at the sites $i$ labeled in Fig. \ref{fig:global_D4with8sites} (a). 
The results indicate that $\langle S^z_i\rangle$ exhibits a linear response behavior when $h_g < 0.03$, which allows us to obtain the actual distribution of $\langle S^z_i\rangle$ with $h_g = 0$ by the linear-fitting method. 
The fitting results after symmetrization with values are shown in Fig. \ref{fig:global_D4with8sites} (d).

The fitting distributions of the magnetic order parameter $\langle S^z_i\rangle$ after symmetrization in other QCs are shown in Fig. \ref{fig:global_other}.

\begin{figure*}[h]
    \centering
    \includegraphics[width=0.6\linewidth]{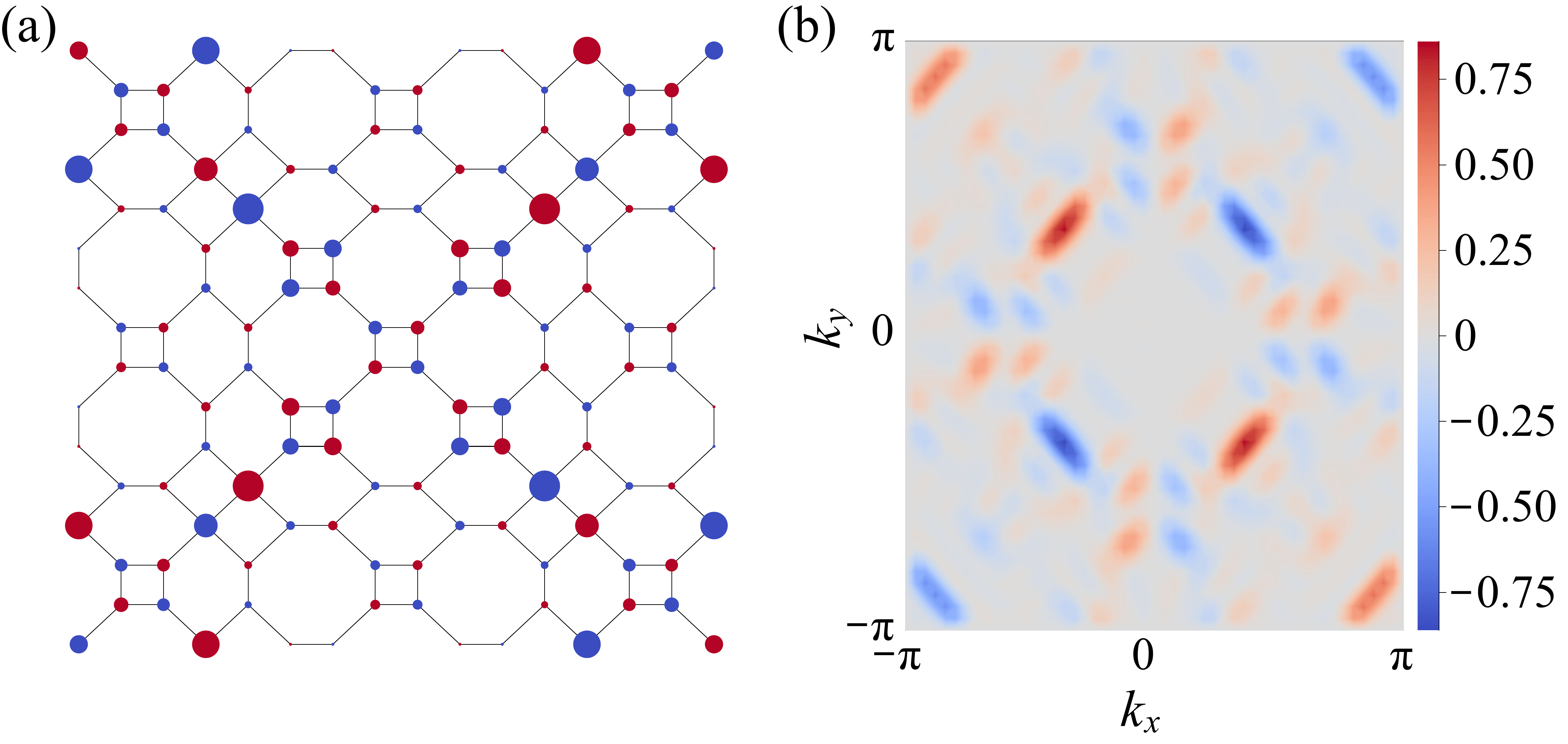}
    \caption{(a) The magnetism distribution in the 128-site $D_4$ symmetric Thue-Morse lattices without the radius restriction.
    (b) The spin-resolved spectral density difference $\mathcal{A}_{\uparrow}(\bm{p},\omega) - \mathcal{A}_{\downarrow}(\bm{p},\omega)$ as a function of momentum $\bm{p}$ for $\omega=-0.2t_1$.}
    \label{fig:ThueMorse_without_r}
\end{figure*}

We also calculate the magnetism distribution in $D_4$ symmetric Thue-Morse lattices without the radius restriction, as shown in Fig. \ref{fig:ThueMorse_without_r} (a). 
Fig. \ref{fig:ThueMorse_without_r} (b) exhibits the corresponding momentum-dependent spin-resolved spectral density difference.
These results in $D_4$ symmetric Thue-Morse lattices without the radius restriction are similar to those in $D_4$ symmetric Thue-Morse lattices with the radius restriction, indicating the robustness of AM order in the structure, which can be explained by symmetry.

\begin{figure*}[h]
    \centering
    \includegraphics[width=1\linewidth]{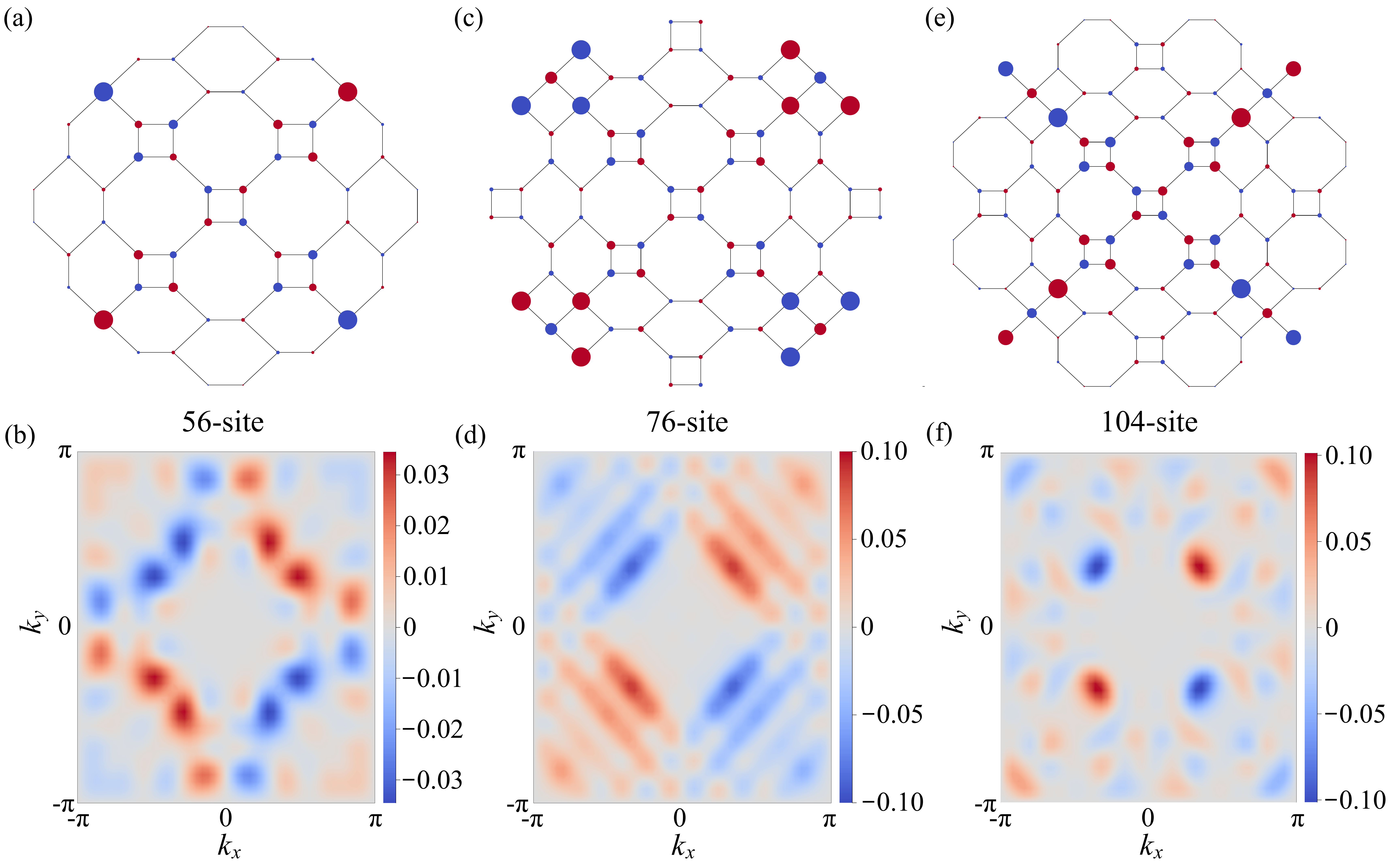}
    \caption{The size dependence of the magnetic properties in the $D_4$-symmetric Thue-Morse lattices.
    The magnetism distributions in $D_4$ symmetric Thue-Morse lattices with 56 sites for (a), 76 sites for (c), and 104 sites for (e).
    (b, d, f) The corresponding spin-resolved spectral density differences $\mathcal{A}_{\uparrow}(\bm{p},\omega) - \mathcal{A}_{\downarrow}(\bm{p},\omega)$ as functions of momentum $\bm{p}$ for the lattices shown in (a), (c), and (e), respectively.}
    \label{figS:ThueMorse_scaling}
\end{figure*}

\begin{figure*}[h]
    \centering
    \includegraphics[width=1\linewidth]{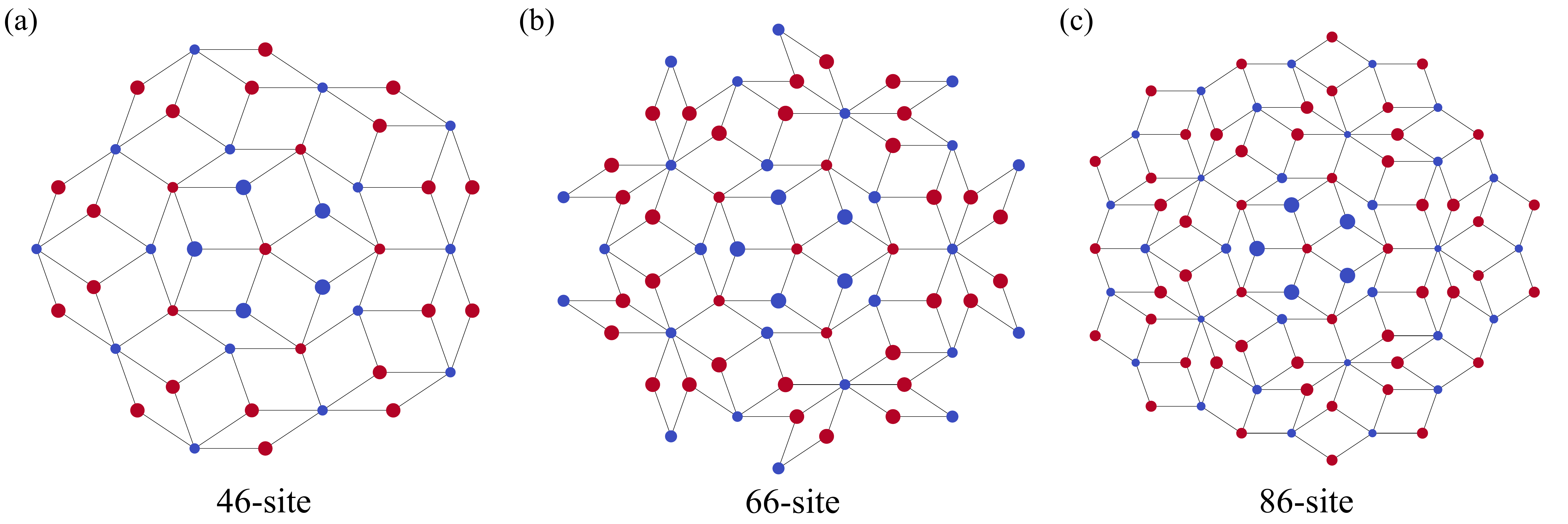}
    \caption{The size dependence of the magnetism distribution in $D_5$-symmetric Penrose lattices with 46 sites for (a), 66 sites for (b), and 86 sites for (c).}
    \label{figS:Penrose_scaling}
\end{figure*}

We further examine the finite-size effects of the N\'eel ordered ground states in several bipartite QCs.
For the $D_4$-symmetric Thue-Morse lattices, Figs. \ref{figS:ThueMorse_scaling} (a), (c), and (e) show that the real-space magnetic pattern keeps the same N\'eel order with opposite spin orientations on the two sublattices for all studied sizes.
The corresponding spin-resolved spectral density differences in Figs. \ref{figS:ThueMorse_scaling} (b), (d), and (f) remain nonzero and momentum-dependent, and their symmetry character is consistent with the real-space AM pattern at different sizes.
For the $D_5$-symmetric Penrose lattices, the magnetic patterns in Figs. \ref{figS:Penrose_scaling} (a-c) also show a stable $s$-wave N\'eel-type order as the system size increases.
Similar size-independent N\'eel ordered patterns are also obtained for the $D_2$-symmetric Thue-Morse and $D_8$-symmetric Ammann-Beenker lattices.
These results show that the N\'eel ordered ground states studied in our work are robust against finite-size effects.

\begin{figure*}[h]
    \centering
    \includegraphics[width=0.45\linewidth]{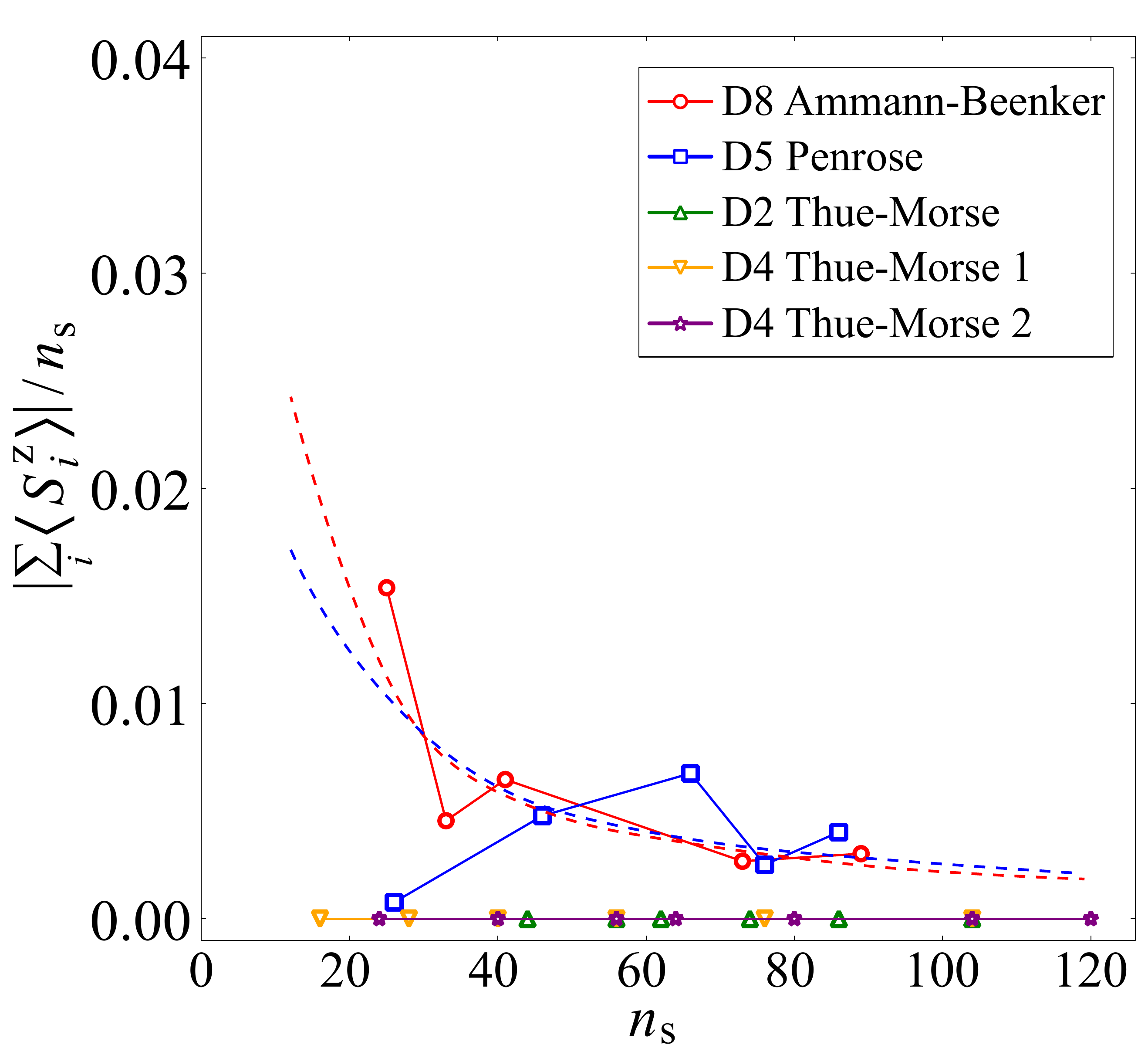}
    \caption{Scaling results for the average net magnetism in different lattices.
    $n_s$ is the number of sites, and the dashed lines are fitting lines.}
    \label{figS:Sz_scaling}
\end{figure*}

We also analyze the size dependence of the average net magnetism, as shown in Fig. \ref{figS:Sz_scaling}.
For the $D_4$-symmetric Thue-Morse lattices, the average net magnetism remains zero at all studied sizes, consistent with the symmetry-protected AM character.
For the $D_5$-symmetric Penrose and $D_8$-symmetric Ammann-Beenker lattices, the finite clusters show nonzero average net magnetism, but its magnitude generally decreases with increasing system size.
This behavior is consistent with the G-AM classification discussed in the main text, where the thermodynamic-limit zero net magnetism is enforced by the special geometry of the quasicrystal rather than by a sublattice-exchanging symmetry.
Therefore, the finite-size analysis supports the robustness of the N\'eel ordered ground states and does not change the magnetic classification in the main text.

% ----- Sec F [end] -----

\end{document}